\documentstyle[epsfig,12pt]{article}
\def\tr{{\rm tr}}
\def\aleq{<\hspace{-4.5mm}\raisebox{-.8ex}{$\sim$}}
\def\sD{{\cal D}\hspace{-2.5mm}/}

\def\spartial{\partial \hspace{-2.2mm}/}

\def\sk{k\hspace{-2mm}/}

\begin{document}

\baselineskip 0.6cm
\begin{titlepage}
\begin{flushright}
TU-516
\end{flushright}
\vskip .5cm
 
\begin{center}
{\LARGE         SUSY and Flat Direction in de Sitter Space }

\vskip 2.0cm

\large{Masahiro Tanaka}
\vskip 1.5cm
{\em Department of Physics, Tohoku University, Sendai, 980-77  Japan}

\end{center}

\begin{center}
{\bf Abstract}
\end{center}
We have found that supersymmetry (SUSY) in curved space is 
broken {\it softly}. It is 
also found that Pauli-Villars regularization preserves the remaining symmetry, 
softly broken SUSY. Using it we computed the one-loop 
effective potential along a (classical) 
flat direction in a Wess-Zumino model 
in de Sitter space. The analysis is 
relevant to the Affleck-Dine mechanism for baryogenesis. 
The effective potential is unbounded from below:
$V_{eff}(\phi)\to 
-3g^2H^2\phi ^2\ln \phi ^2 /16\pi ^2 $,
where $\phi $ is the scalar field along the flat direction, $g$ is a
typical coupling constant, and $H$ is the Hubble parameter.
This is identical with the effective potential which is obtained by using 
proper-time cutoff regularization. 
Since proper-time cutoff regularization is exact even
 at the large curvature region, the effective potential 
possesses softly broken SUSY and reliability in the
large curvature region.
\end{titlepage}

%%%%%%%%%%%%%%%%%%%%%%%%%%%%%%%%%%%%%%%%%%%%%%%%%%%%%%%%%%%%%%%%%%%%%%%%%%%%%%%%%%%%%% INTRODUCTION %%%%%%%%%%%%%%%%%%%%%%%%%%%%%%%%%%%%%%%%%%%%%%%%%%%%%%%%%%%%%%%%%%%%%%%%%%%%%%%%%%%%%%%%%%%%%%%%%%%%%%%%%%%%%%%%%%%%%%%%%%%%%%%%%%%%%%%%%

\section{Introduction}
\label{I}

A scalar field can effectively be considered as massless, if its mass
$m$ is much smaller than the Hubble parameter $H$. 
Such a scalar field has a flat direction in the field space 
along which the potential does not vary.

Our interest in flat directions is motivated by their use
in the Affleck-Dine mechanism~\cite{a&d} for baryogenesis. 
In their scenario, just after inflation the scalar 
field along a flat direction has a large expectation value 
of the order of the GUT (Grand Unified Theory) scale and can be associated 
with baryon number violating operators. 
The large expectation value is due to quantum fluctuation of the scalar 
field during inflation. At the epoch right after inflation the Hubble 
parameter may be of the order of the GUT scale 
or the intermediate scale ($\sim
10^{11}\rm{GeV}$). 
After inflation and when the Hubble 
parameter becomes of the order of the supersymmetry (SUSY) breaking mass 
$m$ (that is of the order of the weak scale $\sim
10^2\rm{GeV}$), the scalar field begins to
fall down along the flat direction toward a true minimum of the potential. 
A condensate of the scalar field (say the squark field) 
possesses a baryon number density which is gradually diluted 
by the expansion of the universe, and 
finally decays into lighter particles. The large initial value\footnote{
The initial conditions are set at the epoch (right) after inflation.}
 mean guarantees 
a large baryon to photon ratio which
could be larger than the observed. In their scenario it is a 
flat direction that gives such a large initial value.
        
In reality the 
Affleck-Dine mechanism works well 
in supersymmetric unified theories because 
a globally supersymmetric model often has 
(exact) flat directions\footnote{If there is no constant term in a potential, 
(exact) flat directions are zero energy states, where SUSY is unbroken.}
 and, what is more important, the (exact) flat directions are 
free from quantum correction 
due to the non-renormalization theorem~\cite{w&b,g&r&s}.
  
 With a soft SUSY breaking mass $m$, on the other hand, 
the potential along the direction receives logarithmically 
divergent radiative correction proportional 
to the soft SUSY breaking mass squared $m^2$. 
But logarithmically divergent 
correction is much milder than quadratically divergent.
If the soft SUSY breaking mass is small, quantum correction is 
also small. Therefore the flat direction is still flat approximately.

Although our motivation for use of flat directions comes from 
cosmology, it is not known if flat directions in curved space 
are (effectively) flat or not
including quantum correction. To understand it
 is the purpose of this paper. 
Hereafter we refer to flat direction at tree (classical)
level as flat direction.   
More specifically, we evaluate the one-loop effective 
potential\cite{c&w,t,o,b&o&s} along
a flat direction in de Sitter space. De Sitter space is of particular interest 
since the inflationary phase of the universe is described by this 
space. Therefore the shape of the potential along 
flat directions in de Sitter space 
is crucial to the Affleck-Dine mechanism. 

In this paper 
we consider {\it renormalizable} 
models that reduce to global (N=1) supersymmetric models in Minkowski space 
limit.\footnote{In Ref.~\cite{o} the one-loop effective
  action of N=1 pure supergravity
  model in de Sitter space is studied.}
 Although we study {\it supersymmetric} models in de Sitter space, 
we do not know how to characterize this 
symmetry because SUSY in curved space is broken.
We must know what kind of SUSY-like symmetry remains in de Sitter space. 
The calculation we perform must respect this symmetry if it is ever 
possible.  
Note that the renormalizability in curved space allows a curvature coupling, 
or the conformal coupling, 
$-\xi {\cal R}\phi ^2$\cite{t,b&o&s,b&d}. 
Since the scalar curvature 
${\cal R}$ is 
given by ${\cal R}=-12H^2$ in de Sitter space, if $\xi$ is of the order of 
unity, the direction is not flat even at tree level.

%%%%%%%%%%%%%%%%%%%%%%%%%%%%%%%%%%%%%%%%%%%%%%%%%%%%%%%%%%%%%%%%%%%%%%%%%%%%%%%%%%%%%%%% SOFTLY BROKEN SUSY IN CURVED SPACE %%%%%%%%%%%%%%%%%%%%%%%%%%%%%%%%%%%%%%%%%%%%%%%%%%%%%%%%%%%%%%%%%%%%%%%%%%%%%%%%%%%%%%%%%%%%%%%%%%%%%%%%%%%%%%%%%

\section{Softly Broken SUSY in Curved Space}
\label{SBSCS}
\subsection{Softly Broken SUSY in Curved Space}
\label{SBSWTI}

We see that global SUSY is broken softly in 
curved space\footnote{In anti-de Sitter space there exists global 
SUSY\cite{b} because of its maximal symmetry, but in de Sitter space
 this is not the fact.}, 
e.g., de Sitter space. A theory with 
softly broken symmetry is characterized 
by the following two conditions\cite{s1}. One is that the new 
divergences which the theory contains other than symmetric one does all 
appear in the operators with dimension smaller than four. The other is that 
the strongest superficial divergence remains unchanged even 
in the theory with softly broken symmetry.
     
For simplicity we will treat the Wess-Zumino model with a superpotential 
\begin{eqnarray}
P=\frac{m}{2}\phi ^2+\frac{\sqrt{2}}{3}g\phi ^3.
\label{11,0}
\end{eqnarray}
In curved space
\begin{eqnarray}
e^{-1}{\cal L}&=&\frac{1}{2}\Bigl(A\Box A
+B\Box B +i\bar{\psi}\sD \psi
+F^2+G^2\Bigr) +\frac{1}{2}\xi {\cal R}\Bigl( A^2+B^2\Bigr) 
\nonumber\\
& &\hspace{-.5cm}+m\biggl( AF-BG+\frac{1}{2}\bar{\psi}\psi \biggr) 
+g \Bigl[ \bar{\psi}(A-\gamma _5B)\psi +F(A^2-B^2)-2GAB\Bigr]  \label{11,1}
\end{eqnarray}   
where $A$ is the real scalar field, $B$ is the real pseudoscalar field
$(\phi =(A+iB)/\sqrt{2})$, and 
$\psi $ is the Majorana spinor field. $F$ and $G$ are the auxiliary fields in 
the scalar (chiral) multiplet. We include the curvature coupling 
with the scalar curvature 
${\cal R}$. $m$ is the supersymmetric mass. 
In flat space limit this Lagrangian is invariant under the 
supertransformation: 
\begin{eqnarray}
&&\delta A=-\bar{\epsilon}\psi ,\hspace{1cm}
\delta B=-\bar{\epsilon}\gamma _5\psi \nonumber\\
&&\delta F=-i\bar{\epsilon}\sD \psi ,\hspace{.6cm}
\delta G=-i\bar{\epsilon}\sD \gamma _5\psi 
\nonumber\\
&&\delta \psi= \Bigl[ i\spartial A+F
-\gamma _5\bigl( i\spartial B+G\bigr) \Bigr] \epsilon \label{11,6},            
\end{eqnarray}
if we choose $\epsilon $ constant. In curved space this is not the fact:
\begin{eqnarray}
e^{-1}\delta {\cal L}&=&{\cal D}_{\mu}\bar{\epsilon}\cdot 
\Bigl[ -\sD \bigl( A-\gamma _5B\bigr) \cdot \gamma ^{\mu}\psi 
+im\bigl( A+\gamma _5B\bigr) \gamma ^{\mu}\psi \Bigr] \nonumber\\
& & -\bar{\epsilon}\xi {\cal R}
\bigl( A+\gamma _5B\bigr) \psi +g{\cal D}_{\mu}\bar{\epsilon}\cdot 
i\bigl(A^2-B^2+2\gamma _5AB\bigr) \gamma ^{\mu}\psi \label{11,7}\\
&\neq &0 .
\end{eqnarray}
To see that SUSY is broken softly we must pick up all the divergent 
1PI diagrams at one-loop order.   
It is done by expressing the propagators in momentum representation 
\cite{b&p}\footnote{We use Riemann normal coordinates in obtaining the 
expressions\cite{m&t&w}.}
  
\begin{eqnarray}
\langle  A(x)A(x')\rangle &=&\langle B(x)B(x')\rangle  
\label{11,8}\\
&=&\frac{i}{\Box +\xi {\cal R}-m^2}\delta (x,x')
\label{11,9}\\
&=&i\int \frac{d^4k}{(2\pi)^4}e^{ik\cdot (x-x')}\Biggl[ -\frac{1}{k^2+m^2}
+\biggl( \frac{1}{6}-\xi \biggr) \frac{{\cal R}}{(k^2+m^2)^2} 
\nonumber\\
& &+\frac{1}{6}\frac{{\cal R}}{(k^2+m^2)^2} 
-\frac{2}{3}\frac{{\cal R}_{ab}k^ak^b}{(k^2+m^2)^3}
+{\cal O}({\cal R}^{3/2})\Biggr] 
\label{11,10}\\
\langle \psi(x)\bar{\psi}(x')\rangle &=&
\frac{i}{i\sD _x +m}\delta (x,x') 
\label{11,17}\\
&=&i\int \frac{d^4k}{(2\pi)^4}e^{ik\cdot (x-x')}
\Biggl[ \frac{\sk +m}{k^2+m^2}
-\frac{{\cal R}}{8}\frac{\sk}{(k^2+m^2)^2} 
\nonumber\\
& &-\frac{{\cal R}}{12}\frac{\sk +m}{(k^2+m^2)^2}
+\frac{2}{3}{\cal R}_{ab}k^ak^b\frac{\sk +m}{(k^2+m^2)^3}
+{\cal O}({\cal R}^{3/2})\Biggr] 
\label{11,18}\\
\langle A(x)F(x')\rangle &=&-\langle B(x)G(x')\rangle =
-m\langle  A(x)A(x')\rangle \label{11,11}\\
\langle F(x)F(x')\rangle &=&\langle G(x)G(x')\rangle =
i\delta (x,x')+m^2\langle  A(x)A(x')\rangle 
\label{11,14}.
\end{eqnarray}

We can find that one, two, and three point vertex functions are 
logarithmically divergent and divergence due to the 
curvature appears only in one and two point vertex functions. 
Superficial degrees of divergence is unchanged in each
vertex function. From the above argument it can be deduced that 
the curvature acts as the soft SUSY breaking mass.  See Table~\ref{2,hyou}.

\begin{table}[h]
\begin{center}
\renewcommand{\arraystretch}{1.4}  
\begin{tabular}{|c|lll|lll|}         \hline
Dim. of  & \multicolumn{6}{c|}{Degrees of divergence}\\\cline{2-7}
operator & \multicolumn{3}{c|}{SUSY} &\multicolumn{3}{c|}{ Broken
  SUSY in curved space}  \\ \hline 
 1  & cubic & $\to $& nothing & cubic& $\to $ &logarithmic \\
 2  & quadratic& $\to $& logarithmic & quadratic& $\to $& logarithmic \\
 3  & linear& $\to $& logarithmic &  linear& $\to $& logarithmic \\
 4  & logarithmic& $\to $& logarithmic & logarithmic& $\to $& logarithmic \\
\hline
\end{tabular}
\caption{This table shows the structure of divergence appear in the
  coefficients of operators. The left hand side of each arrow is the 
superficial degrees of divergence and the right hand side is its 
actual degrees of divergence.}
\label{2,hyou}
\end{center}
\end{table}

The reason why we can get such observations is as follows. 
One-loop diagrams which contribute to the one point function cancel 
one another in exactly supersymmetric theories 
while each of them diverges quadratically. In de Sitter space (or in
curved space) the 
curvature dependent part breaks cancellation but it just gives 
logarithmic divergence because a propagator is expanded by power of 
${\cal R}\times ({\rm
  momentum})^{-2}$ ( see Eqs.~(\ref{11,10})
and (\ref{11,18})). One-loop 
contribution to the two point vertex functions is non-vanishing even in 
exactly supersymmetric theories and gives logarithmic divergence. 
Moreover, one-loop contribution from the curvature dependent parts
exists and is 
logarithmically divergent just like the case of one point function.   
The three point function diverges logarithmically even 
at each graph, so the curvature dependent parts are finite.
A reason from another point of view is explained in Appendix~\ref{CTWZ}.

In this paper we consider that all the suitable regularizations for SUSY 
in curved space must satisfy the Ward-Takahashi identity 
for softly broken SUSY.

%%%%%%%%%%%%%%%% WTI for SBS with PVR %%%%%%%%%%%%%%%%%%%%%%%%%%%%%%%%
%%%%%%%%%%%%%%%%%%%%%%%%%%%%%%%%%%%%%%%%%%%%%%%%%%%%%%%%%%%%%%%%%%%%
\subsection{Ward-Takahashi Identity for Softly Broken SUSY with
  Pauli-Villars Regulators}
 
We derive the Ward-Takahashi identity for softly broken SUSY. 
We adopt Pauli-Villars regularization as a candidate of suitable 
regularization for softly broken 
SUSY in de Sitter space. It is manifestly supersymmetric in exactly 
supersymmetric theories\cite{g&g&r&s,b}.
To check if Pauli-Villars regularization satisfies 
the Ward-Takahashi 
identity or not we write down the total Lagrangian with the regulators.
The superpotential with the regulators is as follows
\begin{eqnarray}
P=\frac{1}{2}\sum _im_i{\phi _i}^2+\frac{\sqrt{2}}{3}\sum _{ijk}
g_{ijk}\phi_i\phi _j\phi _k .
\label{ssp}
\end{eqnarray}
To regularize all the divergent diagrams the regulator fields have the
same coupling constant with the physical field:
\begin{eqnarray}
g_{ijk}=g.
\end{eqnarray}
We obtain the Lagrangian;
\begin{eqnarray}
e^{-1}{\cal L}_{tot}&=&\sum _i\frac{1}{2c_i}\Bigl[ A_i\Box A_i
+B_i\Box B_i 
+i\bar{\psi}_i\sD \psi _i +{F_i}^2+{G_i}^2 
+\xi {\cal R}\Bigl( {A_i}^2+{B_i}^2\Bigr) \nonumber\\
& &+m_i\bigl( 2A_iF_i-2B_iG_i+\bar{\psi}_i\psi _i 
\bigr) \Bigr] +g\sum _{ijk}\Bigl[ \bar{\psi}_i(A_j-\gamma _5B_j)\psi _k
+F_i(A_jA_k \nonumber\\
& &-B_jB_k)
-2G_iA_jB_k\Bigr] 
+\sum _i\Bigl[ J_A^iA_i+J_B^iB_i+J_F^iF_i+J_G^iG_i+\bar{\psi}_i\eta ^i
\Bigr] \nonumber\\
& &+\sum _i\frac{1}{c_i}
\Bigl[ \bar{K}_{\partial A}^{\nu}\partial _{\nu}A_i
+\bar{K}_{\partial B}^{\nu}\partial _{\nu}B_i
+m_i(\bar{K}_AA_i+\bar{K}_BB_i)
+\xi (\bar{K'}_AA_i\nonumber\\
& &+\bar{K'}_BB_i) 
\Bigr] \psi _i +g\sum _{ijk}\Bigl[ \bar{K}_{A^2}A_iA_j
+\bar{K}_{B^2}B_iB_j
+2\bar{K}_{AB}A_iB_j\Bigr] \psi _k.\label{11,19}
\end{eqnarray} 
$J$s are the sources of the bosonic fields and $\eta $ is the source of
the fermionic field. $\bar{K}$s are the sources of the 
composite operators 
which appear in the first variation of the Lagrangian Eq.(\ref{11,7}).
To regularize all the divergent terms we need the Pauli-Villars 
constraints $\sum _ic_i{m_i}^p=0$ for $p=0,1,2$ where $c_0=1$ and
$m_0=m$.
 The regulators, which have $i=1,2,\cdots $, have 
adjustable masses which we set infinite after all the
calculations. $c_i$ is the function of $m_i$s.\footnote{We can get
  the explicit form of $c_i$ as follows. $c_i=-(m_j-m)(m_k-m)/
(m_j-m_i)(m_k-m_i)$ for $i\neq j$, $j\neq k$ and $k\neq i$. 
If we set $m_1<m_2<m_3$, then 
$c_1,c_3 <0$ and $c_2>0$. But we do not need the explicit forms of
$c_i$s.} 
The supertransformation of the total action is
\begin{eqnarray}
\delta S_{tot}&=&\int d^4xe\biggl[ 
\sum _i\frac{1}{c_i}{\cal D}_{\mu}\bar{\epsilon}\cdot 
\Bigl\{ -\sD (A_i-\gamma _5B_i)\gamma ^{\mu}\psi _i
+im(A_i+\gamma_5B_i)\gamma ^{\mu}\psi _i\Bigr\}\nonumber\\
& &-\sum _i\frac{1}{c_i}\bar{\epsilon}\xi {\cal R}
\bigl( A_i +B_i\gamma _5\bigr)\psi _i 
+g\sum _{ijk}{\cal D}_{\mu}\bar{\epsilon}\cdot i\bigl(A_iA_j-B_iB_j
+2\gamma _5A_iB_j\bigr) \gamma ^{\mu}\psi _k  \nonumber\\
& &-\bar{\epsilon}\sum _i\Bigl\{ \psi _iJ_A^i+\gamma _5\psi _iJ_B^i
+i\sD \psi _i\cdot J_F^i
+i\sD \gamma _5\psi _i\cdot J_G^i
\nonumber\\
& &+\bigl(i\spartial A_i-F_i
-i\spartial B_i\gamma _5+G_i\gamma _5\bigr)\eta ^i\Bigr\}+
(\mbox{$\bar{K}$-terms}) \biggr] .\label{11,20}
\end{eqnarray}
It is translated to Ward-Takahashi identity by using generating 
functional
$Z[J,\eta,\bar{K}]$ as follows
\begin{eqnarray}
& &\int d^4xe\biggl[ -{\cal D}_{\mu}\bar{\epsilon}\cdot \biggl\{ 
\gamma ^{\nu}\gamma ^{\mu}\biggl( \frac{\delta}{\delta 
\bar{K}_{\partial A}^{\nu}}
+\gamma _5\frac{\delta}{\delta \bar{K}_{\partial B}^{\nu}}
\biggr) 
-i\gamma ^{\mu}\biggl( \frac{\delta}{\delta \bar{K}_A}
-\gamma _5\frac{\delta}{\delta \bar{K}_B}\biggr) \biggr\}
\nonumber\\
& &-\bar{\epsilon}{\cal R}\biggl( \frac{\delta}{\delta \bar{K'}_A}
+\gamma _5\frac{\delta}{\delta \bar{K'}_B}\biggr)
+{\cal D}_{\mu}\bar{\epsilon}\cdot i\gamma ^{\mu}\biggl( 
\frac{\delta}{\delta \bar{K}_{A^2}}
-\frac{\delta}{\delta \bar{K}_{B^2}} 
-\gamma _5\frac{\delta}{\delta \bar{K}_{AB}}\biggr) \nonumber\\
& &-\bar{\epsilon}\sum _i\biggl\{ J_A^i\frac{\delta}{\delta \bar{\eta}^i}
+J_B^i\gamma _5\frac{\delta}{\delta \bar{\eta}^i}
+iJ_F^i\sD \frac{\delta}{\delta \bar{\eta}^i}
+iJ_G^i\sD \gamma _5\frac{\delta}{\delta \bar{\eta}^i}
+i\gamma ^{\mu}\eta ^i\partial _{\mu}\frac{\delta}{\delta J_A^i} \nonumber\\
& &-\eta ^i\frac{\delta}{\delta J_F^i}
-i\gamma ^{\mu}\gamma _5\eta ^i\partial _{\mu}\frac{\delta}{\delta J_B^i}
+\gamma _5\eta ^i\frac{\delta }{\delta J_G^i}\biggr\} 
+(\mbox{$\bar{K}$-terms})\biggr] Z[J,\eta ,\bar{K}] =0 \label{11,21}.
\end{eqnarray}
Using the connected generating functional $W=-i\ln {Z}$ 
we define the effective action as 
follows
\begin{eqnarray}
\Gamma[\phi ,\psi, \bar{K}]=W[J,\eta,\bar{K}]-\int d^4xe
\Bigl( J\cdot \phi + \bar{\psi}\cdot \eta \Bigr) \label{11,22} .
\end{eqnarray}
Here we denote the bosonic classical fields by $\phi$ and the fermionic
one by $\psi$.  Then 
\begin{eqnarray}
\phi (x)=\frac{\delta}{\delta J(x)}W[J,\eta,\bar{K}],
& &J(x)=-\frac{\delta}{\delta \phi (x)}\Gamma [\phi ,\psi, \bar{K}]
\label{11,23}\\
\psi (x)=\frac{\delta}{\delta \bar{\eta}(x)}W[J,\eta,\bar{K}] ,
& &\eta (x)=-\frac{\delta}{\delta \bar{\psi}(x)}\Gamma [\phi ,\psi, \bar{K}]
\label{11,24}\\
{\cal Q}(x)=\frac{\delta}{\delta \bar{K}(x)}W[J,\eta,\bar{K}]
 &=&\frac{\delta}{\delta \bar{K}(x)}\Gamma [\phi ,\psi, \bar{K}] 
\label{11,26} ,
\end{eqnarray}
where ${\cal Q}(x)$ represents every composite operator.
We obtain the Ward-Takahashi identity for the effective action 
\begin{eqnarray}
& &\int d^4xe\biggl[ -{\cal D}_{\mu}\bar{\epsilon}\cdot \biggl\{ 
\gamma ^{\nu}\gamma ^{\mu}\biggl( 
\frac{\delta \Gamma}{\delta \bar{K}_{\partial A}^{\nu}}
+\gamma _5\frac{\delta \Gamma}{\delta \bar{K}_{\partial B}^{\nu}}\biggr) 
-i\gamma ^{\mu}\biggl( \frac{\delta \Gamma}{\delta \bar{K}_A}
-\gamma _5\frac{\delta \Gamma}{\delta \bar{K}_B}\biggr) \biggr\}
\nonumber\\
& &-\bar{\epsilon}{\cal R}\biggl( \frac{\delta \Gamma}{\delta \bar{K'}_A}
+\gamma _5\frac{\delta \Gamma}{\delta \bar{K'}_B}\biggr)
+{\cal D}_{\mu}\bar{\epsilon}\cdot i\gamma ^{\mu}\biggl( 
\frac{\delta \Gamma}{\delta \bar{K}_{A^2}}
-\frac{\delta \Gamma}{\delta \bar{K}_{B^2}} 
-\gamma _5\frac{\delta \Gamma}{\delta \bar{K}_{AB}}\biggr) \nonumber\\
& &+\bar{\epsilon}\sum _i\biggl\{ \frac{\delta \Gamma}{\delta A_i}\psi _i
+\frac{\delta \Gamma}{\delta B_i}\gamma _5\psi _i
+i\frac{\delta \Gamma}{\delta F_i}\sD \psi _i
+i\frac{\delta \Gamma}{\delta G_i}\sD \gamma _5\psi _i 
+i\gamma ^{\mu}\frac{\delta \Gamma}{\delta \bar{\psi}_i}\partial _{\mu}A_i 
\nonumber\\
& &-\frac{\delta \Gamma}{\delta \bar{\psi}_i}F_i
-i\gamma ^{\mu}\gamma _5\frac{\delta \Gamma}{\delta \bar{\psi}_i}
\partial _{\mu}B_i
+\gamma _5\frac{\delta \Gamma}{\delta \bar{\psi}_i}G_i \biggr\} 
+(\mbox{$\bar{K}$-terms})\biggr] =0. \label{11,27}
\end{eqnarray}
All the fields in effective actions are classical fields, although we do 
not make no distinction between a classical field and 
a quantum one in notations.
In the previous subsection we found that the divergent parts of 
the effective action are those 
of one, two, and three point vertex functions. Three 
point function diverges only at the curvature independent part, 
supersymmetric one, so we do not need to check the Ward-Takahashi identity 
for three point functions. 
We do not need the $\bar{K}$-terms explicitly, because in deriving 
the Ward-Takahashi identity for n-point vertex function we do not 
differentiate by $\bar{K}$ and do set them zero after the differentiations
with respect to fields.

\subsection{One Point Function}
\label{OPF}

In this section we check the Ward-Takahashi identity for one point vertex 
function.    
We differentiate Eq.($\ref{11,27}$) with respect to
 $\psi _j(y)$ and set all the 
fields and the sources of composite fields zero. We obtain 
\begin{eqnarray}
& &\int d^4xe\biggl[ -{\cal D}_{\mu}\bar{\epsilon}\cdot \biggl\{ 
\gamma ^{\nu}\gamma ^{\mu}\biggl( 
\frac{\delta ^2\Gamma}
{\delta \psi _j(y)\delta \bar{K}_{\partial A}^{\nu}}
+\gamma _5\frac{\delta ^2\Gamma}
{\delta \psi _j(y)\delta \bar{K}_{\partial B}^{\nu}}\biggr) 
-i\gamma ^{\mu}\biggl( \frac{\delta ^2\Gamma}
{\delta \psi _j(y)\delta \bar{K}_A}\nonumber\\
& &-\gamma _5\frac{\delta ^2\Gamma}
{\delta \psi _j(y)\delta \bar{K}_B}\biggr) \biggr\}
-\bar{\epsilon}{\cal R}\biggl( \frac{\delta ^2\Gamma}
{\delta \psi _j(y)\delta \bar{K'}_A}
+\gamma _5\frac{\delta ^2\Gamma}
{\delta \psi _j(y)\delta \bar{K'}_B}\biggr)
+{\cal D}_{\mu}\bar{\epsilon}\cdot i\gamma ^{\mu}\cdot  
\nonumber\\ 
&&\biggl( \frac{\delta ^2\Gamma}{\delta \psi _j(y)\delta \bar{K}_{A^2}}
-\frac{\delta ^2\Gamma}{\delta \psi _j(y)\delta \bar{K}_{B^2}} 
-\gamma _5\frac{\delta ^2\Gamma}
{\delta \psi _j(y)\delta \bar{K}_{AB}}\biggr) 
+\bar{\epsilon}\frac{\delta \Gamma}{\delta A_j}\delta (x,y) \biggr] =0. 
\label{11,28}
\end{eqnarray}
This is the non-trivial Ward-Takahashi identity for one point vertex function.
Then we calculate the all the terms of the left hand side of Eq.($\ref{11,28}$)
at one-loop level.
We begin with the explicit computation of the first term in 
Eq.($\ref{11,28}$);
\begin{eqnarray}
\frac{\delta ^2W_1}
{\delta \eta ^{j'}(y')\delta \bar{K}^{\nu}_{\partial A}(x)}
&=&-\int d^4u
\langle \bar{\psi}_{j'}(y')
\sum _i\frac{1}{c_i}\partial _{\nu x}A_i(x)\cdot \psi _i(x)
{\cal L}_{int}(u)\rangle 
\nonumber\\
&=&2g\int d^4ue_u\langle \psi _{j'}(u)\bar{\psi}_{j'}(y')\rangle 
\sum _i\frac{1}{c_i}\langle \psi _i(x)\bar{\psi}_i(u)\rangle 
\nonumber\\
& &\hspace{1cm}\cdot \langle \partial _{\nu x}A_i(x)A_i(u)\rangle .
\label{11,29}
\end{eqnarray}

We obtain the first term in Eq.($\ref{11,28}$)
\begin{eqnarray}
\frac{\delta ^2\Gamma_1}
{\delta \psi _j(y)\delta \bar{K}^{\nu}_{\partial A}(x)}
&=&-2ig\sum _i\frac{1}{c_i}\langle \psi _i(x)\bar{\psi }_i(y)\rangle 
\langle \partial _{\nu x}A_i(x)A_i(y)\rangle  
\nonumber\\
&=&2ig\sum _i\frac{c_i}{i\sD _x+m_i}\delta (x,y)
\cdot \frac{\partial _{\nu x}}{\Box _x+\xi {\cal R}-{m_i}^2}\delta (x,y).
\label{11,31}
\end{eqnarray}

\begin{figure}[ht]
\begin{center}
\epsfig{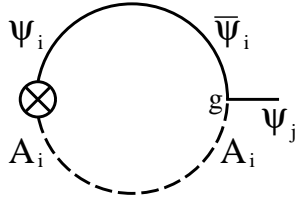}
\caption{One Point Function $\langle \psi \rangle _{tr}$ (i)}
\label{11,a}
\end{center}
\end{figure}

This diagram (Figure~\ref{11,a}) is quadratically divergent. But it does 
not contribute to the effective action at all 
because we set all the source terms zero.

Also for the second term, via the connected two point function:  
\begin{eqnarray}
\frac{\delta ^2W_1}
{\delta \eta ^{j'}(y')\delta \bar{K}^{\nu}_{\partial B}(x)}
&=&-\int d^4u
\langle \bar{\psi}_{j'}(y')
\sum _i\frac{1}{c_i}\partial _{\nu x}B_i(x)\cdot \psi _i(x)
{\cal L}_{int}(u)\rangle 
\nonumber\\
&=&-2g\int d^4ue_u\langle \psi _{j'}(u)\bar{\psi}_{j'}(y')\rangle 
\sum _i\frac{1}{c_i}\langle \psi _i(x)\bar{\psi}_i(u)\rangle 
\nonumber\\
& &\hspace{1cm}\cdot \gamma _5\langle \partial _{\nu x}B_i(x)B_i(u)
\rangle ,
\label{11,32}
\end{eqnarray}
and we obtain (Figure~\ref{11,b}) 
\begin{eqnarray}
\frac{\delta ^2\Gamma_1}
{\delta \psi _j(y)\delta \bar{K}^{\nu}_{\partial B}(x)}
&=&2ig\sum _i\frac{1}{c_i}\langle \psi _i(x)\bar{\psi }_i(y)\rangle \gamma _5
\langle \partial _{\nu x}B_i(x)B_i(y)\rangle  \nonumber\\
&=&-2ig\sum _i\frac{c_i}{i\sD _x+m_i}\delta (x,y)\gamma _5
\cdot \frac{\partial _{\nu x}}{\Box _x+\xi {\cal R}-{m_i}^2}\delta (x,y).
\label{11,34}
\end{eqnarray}

\begin{figure}[ht]
\begin{center}
\epsfig{file=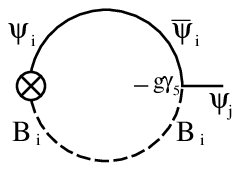,width=4cm}
\caption{One Point Function $\langle \psi \rangle _{tr}$ (ii)}
\label{11,b}
\end{center}
\end{figure}
Likewise we obtain 
\begin{eqnarray}
\frac{\delta ^2\Gamma_1}
{\delta \psi _j(y)\delta \bar{K}_A(x)}
&=&-2ig\sum _i\frac{m_i}{c_i}
\langle \psi _i(x)\bar{\psi }_i(y)\rangle \langle A_i(x)A_i(y)\rangle 
\nonumber\\
&=&2ig\sum _i\frac{c_im_i}{i\sD _x+m_i}\delta (x,y)
\cdot \frac{1}{\Box _x+\xi {\cal R}-{m_i}^2}\delta (x,y),
\label{11,36}\\
\frac{\delta ^2\Gamma_1}
{\delta \psi _j(y)\delta \bar{K}_B(x)}
&=&2ig\sum _i\frac{m_i}{c_i}\langle \psi _i(x)\bar{\psi }_i(y)\rangle 
\gamma _5\langle B_i(x)B_i(y)\rangle 
\nonumber\\
&=&-2ig\sum _i\frac{c_im_i}{i\sD _x+m_i}\delta (x,y)\gamma_5
\cdot \frac{1}{\Box _x+\xi {\cal R}-{m_i}^2}\delta (x,y),
\label{11,38}\\
\frac{\delta ^2\Gamma_1}
{\delta \psi _j(y)\delta \bar{K'}_A(x)}
&=&-2ig\sum _i\frac{\xi}{c_i}\langle \psi _i(x)\bar{\psi }_i(y)\rangle 
\langle A_i(x)A_i(y)\rangle 
\nonumber\\
&=&2ig\sum _i\frac{c_i\xi}{i\sD _x+m_i}\delta (x,y)
\cdot \frac{1}{\Box _x+\xi {\cal R}-{m_i}^2}\delta (x,y),
\label{11,40}
\end{eqnarray}
and 
\begin{eqnarray}
\frac{\delta ^2\Gamma_1}
{\delta \psi _j(y)\delta \bar{K'}_B(x)}
&=&2ig\sum _i\frac{\xi}{c_i}\langle \psi _i(x)\bar{\psi }_i(y)\rangle 
\gamma _5\langle B_i(x)B_i(y)\rangle 
\nonumber\\
&=&-2ig\sum _i \frac{c_i\xi}{i\sD _x+m_i}\delta (x,y)\gamma _5
\cdot \frac{1}{\Box _x+\xi {\cal R}-{m_i}^2}\delta (x,y).
\label{11,42}
\end{eqnarray}

Finally we calculate the one point vertex function of the real scalar
$A$.
One point function is
\begin{eqnarray}
\frac{\delta W_1}{\delta J_A^{j'}(y')}
&=&i\int d^4u\langle A_{j'}(y'){\cal L}_{int}(u)\rangle 
\nonumber\\
&=&ig\int d^4ue_u\sum _i\langle A_{j'}(y')A_{j'}(u)\rangle 
\Bigl[ -\tr{\langle \psi _i(u)\bar{\psi}_i(u)\rangle }
\nonumber\\
& &\hspace{1cm}+2\langle F_i(u)A_i(u)\rangle 
-2\langle G_i(u)B_i(u)\rangle \Bigr] .
\label{11,43}
\end{eqnarray}
We obtain the non-vanishing one point vertex function (Figure~\ref{11,c})
\begin{eqnarray}
\frac{\delta \Gamma _1}{\delta A_j(y)}&=&g\sum _i
\Bigl[ \tr{\langle \psi _i(y)\bar{\psi}_i(y)\rangle }
-2\langle F_i(y)A_i(y)\rangle +2\langle G_i(y)B_i(y)\rangle \Bigr] 
\nonumber\\
&=&ig\sum _ic_i\Biggl[ \tr{\frac{1}{i\sD _y+m_i}}
+\frac{4m_i}{\Box _y+\xi {\cal R}-{m_i}^2}\Biggr] \delta (y,y).
\label{11,45}
\end{eqnarray}

\begin{figure}[ht]
\begin{center}
\epsfig{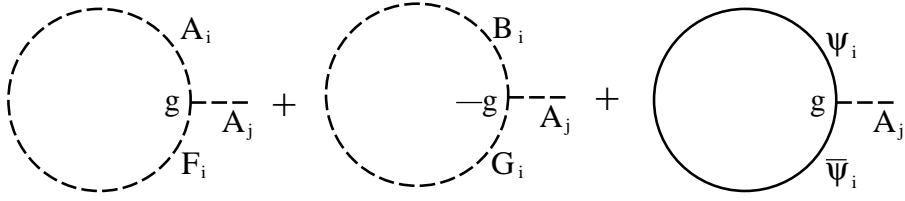}
\caption{One Point Function $\langle A\rangle _{tr}$}
\label{11,c}
\end{center}
\end{figure}

Inserting all of them into left hand side of 
the Ward-Takahashi identity for one point 
function Eq.$(\ref{11,28})$ we can see that it is satisfied.\footnote{ The 
inclusion of linear $A_i$ term in the potential 
amounts to a trivial relation like Eq.(\ref{11,23}) at tree level}
Note that the third line in Eq.$(\ref{11,28})$ is zero itself.

\subsection{Two Point Function}
\label{TPF}

In this subsection we check the Ward-Takahashi identity for two point vertex 
function. We explicitly 
check one of them, Ward-Takahashi identity between real scalar and spinor two
point functions because others are done in the same way.      
To do that we differentiate Eq.($\ref{11,27}$) 
with respect to $\psi _j(y)$ and $A_k(z)$ and set all the 
fields and the sources of composite fields zero. We obtain
\begin{eqnarray}
& &\int d^4xe\biggl[ -{\cal D}_{\mu}\bar{\epsilon}\cdot \biggl\{ 
\gamma ^{\nu}\gamma ^{\mu}\biggl( 
\frac{\delta ^3\Gamma}
{\delta A_k(z)\delta \psi _j(y)\delta \bar{K}_{\partial A}^{\nu}}
+\gamma _5\frac{\delta ^3\Gamma}
{\delta A_k(z)\delta \psi _j(y)\delta \bar{K}_{\partial B}^{\nu}}
\biggr) 
\nonumber\\
& &-i\gamma ^{\mu}\biggl( \frac{\delta ^3\Gamma}
{\delta A_k(z)\delta \psi _j(y)\delta \bar{K}_A}
-\gamma _5\frac{\delta ^3\Gamma}
{\delta A_k(z)\delta \psi _j(y)\delta \bar{K}_B}\biggr) \biggr\}
\nonumber\\
& &-\bar{\epsilon}{\cal R}\biggl( \frac{\delta ^3\Gamma}
{\delta A_k(z)\delta \psi _j(y)\delta \bar{K'}_A}
+\gamma _5\frac{\delta ^3\Gamma}
{\delta A_k(z)\delta \psi _j(y)\delta \bar{K'}_B}\biggr)
+{\cal D}_{\mu}\bar{\epsilon}\cdot i\gamma ^{\mu}\biggl( 
\nonumber\\
& &\frac{\delta ^3\Gamma}
{\delta A_k(z)\delta \psi _j(y)\delta \bar{K}_{A^2}}
-\frac{\delta ^3\Gamma}
{\delta A_k(z)\delta \psi _j(y)\delta \bar{K}_{B^2}} 
-\gamma _5\frac{\delta ^3\Gamma}
{\delta A_k(z)\delta \psi _j(y)\delta \bar{K}_{AB}}\biggr) 
\nonumber\\
& &+\bar{\epsilon}\biggl\{ \frac{\delta ^2\Gamma}
{\delta A_k(z)\delta A_j}\delta (x,y) 
+i\gamma ^{\mu}\frac{\delta ^2\Gamma}{\delta \psi _j(y)\delta \bar{\psi}_k}
\partial _{\mu}\delta (x,z)\biggr\} \biggr] =0. 
\label{11,46}
\end{eqnarray}

The first term is obtained from a two point function:
\begin{eqnarray}
& &\frac{\delta ^3W_1}
{\delta J_A^{k'}(z')\delta \eta ^{j'}(y')
\delta \bar{K}_{\partial A}^{\nu}(x)}
\nonumber\\
&=&\frac{1}{2}\int d^4ud^4v\sum _i\frac{1}{c_i}
\langle A_{k_1}(z_1)\bar{\psi}_{j_1}(y_1)\partial _{\nu x}A_i(x)\cdot 
\psi _i(x){\cal L}_{int}(u){\cal L}_{int}(v)\rangle 
\nonumber\\
&=&-4g^2\int d^4ud^4ve_ue_v\sum _{il}\frac{1}{c_i}\Bigl[ 
\langle A_{k'}(z')A_{k'}(u)\rangle 
\langle \psi _{j'}(v)\bar{\psi}_{j'}(y')\rangle 
\nonumber\\
& &\cdot \langle \partial _{\nu x}A_i(x)A_i(v)\rangle 
\langle \psi _i(x)\bar{\psi}_i(u)\rangle 
\langle \psi _l(x)\bar{\psi}_l(v)\rangle 
+\langle A_{k'}(z')A_{k'}(v)\rangle 
\nonumber\\ 
& &\cdot \langle \psi _{j'}(u)\bar{\psi}_{j'}(y')\rangle 
\Bigl\{ \langle \partial _{\nu x}A_i(x)A_i(v)\rangle 
\langle A_l(u)F_l(v)\rangle  
+\langle \partial _{\nu x}A_i(x)F_i(v)\rangle 
\nonumber\\ 
& &\cdot \langle A_l(u)A_l(v)\rangle \Bigr\} 
\langle \psi _i(x)\bar{\psi }_i(u)\rangle +\cdots \Bigr] . 
\label{11,47}
\end{eqnarray}
The irrelevant part is omitted in the above expression. 
From the above equation we can read off the corresponding vertex function 
(Figure~\ref{11,d})
\begin{eqnarray}
& &\frac{\delta ^3\Gamma _1}
{\delta A_k(z)\delta \psi _j(y)\delta \bar{K}_{\partial A}^{\nu}(x)}
\nonumber\\
&=&4g^2\sum _{il}\frac{1}{c_i}\Bigl[ 
\langle \partial _{\nu x}A_i(x)A_i(y)\rangle 
\langle \psi _i(x)\bar{\psi}_i(z)\rangle 
\langle \psi _l(z)\bar{\psi}_l(y)\rangle 
+\Bigl\{ \langle \partial _{\nu x}A_i(x)A_i(z)\rangle 
\nonumber\\
& &\cdot \langle A_l(y)F_l(z)\rangle 
+\langle \partial _{\nu x}A_i(x)F_i(z)\rangle \langle
A_l(y)A_l(z)\rangle 
\Bigr\} 
\langle \psi_i(x)\bar{\psi}_i(y)\rangle \Bigr] 
\nonumber\\
&=&-4ig^2\sum _{il}c_ic_l\Biggl[ 
\frac{\partial _{\nu x}}{\Box _x+\xi {\cal R}-{m_i}^2}\delta (x,y)
\cdot \frac{1}{i\sD _x+m_i}\delta (x,z)
\nonumber\\
& & \cdot \frac{1}{i\sD _z+m_l}\delta (z,y) 
-\Biggl\{\frac{\partial _{\nu x}}{\Box _x+\xi {\cal R}-{m_i}^2}\delta (x,z) 
\cdot \frac{m_l}{\Box _y+\xi {\cal R}-{m_l}^2}\delta (y,z)
\nonumber\\
& &+\frac{m_i\partial _{\nu x}}{\Box _x+\xi {\cal R}-{m_i}^2}\delta (x,z)
\cdot \frac{1}{\Box _y+\xi {\cal R}-{m_l}^2}\delta (y,z)\Biggr\} 
\frac{1}{i\sD _x+m_i}\delta (x,y)\Biggr] ,
\label{11,49}
\end{eqnarray} 

\begin{figure}[ht]
\begin{center}
\epsfig{file=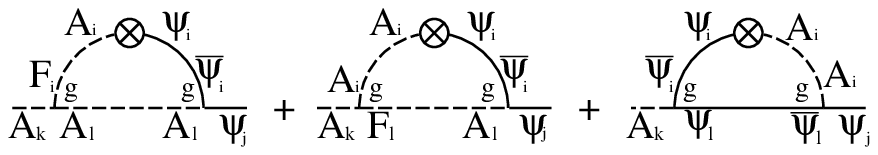,width=13cm}
\caption{Two Point Function $\langle \psi A\rangle _{tr}$(i)}
\label{11,d}
\end{center}
\end{figure}

which is linearly divergent.
In the same way the two point vertex functions with composite operators are 
calculated (Figure~\ref{11,d} and \ref{11,e}):
\begin{eqnarray}
& &\frac{\delta ^3\Gamma _1}
{\delta A_k(z)\delta \psi _j(y)\delta \bar{K}_{\partial B}^{\nu}(x)}
\nonumber\\
&=&-4g^2\sum _{il}\frac{1}{c_i}\Bigl[ 
\langle \partial _{\nu x}B_i(x)B_i(y)\rangle 
\langle \psi _i(x)\bar{\psi}_i(z)\rangle 
\langle \psi _l(z)\bar{\psi}_l(y)\rangle 
-\Bigl\{ \langle \partial _{\nu x}B_i(x)B_i(z)\rangle
\nonumber\\
& &\cdot \langle B_l(y)G_l(z)\rangle 
+\langle \partial _{\nu x}B_i(x)G_i(z)\rangle 
\langle B_l(y)B_l(z)\rangle \Bigr\} 
\langle \psi_i(x)\bar{\psi}_i(y)\rangle \Bigr] \gamma _5
\nonumber\\
&=&4ig^2\sum _{il}c_ic_l\Biggl[ 
\frac{\partial _{\nu x}}{\Box _x+\xi {\cal R}-{m_i}^2}\delta (x,y)
\cdot \frac{1}{i\sD _x+m_i}\delta (x,z)
\nonumber\\
& & \cdot \frac{1}{i\sD _z+m_l}\delta (z,y) 
-\Biggl\{\frac{\partial _{\nu x}}{\Box _x+\xi {\cal R}-{m_i}^2}\delta (x,z) 
\cdot \frac{m_l}{\Box _y+\xi {\cal R}-{m_l}^2}\delta (y,z)
\nonumber\\
& &+\frac{m_i\partial _{\nu x}}{\Box _x+\xi {\cal R}-{m_i}^2}\delta (x,z)
\cdot \frac{1}{\Box _y+\xi {\cal R}-{m_l}^2}\delta (y,z)\Biggr\} 
\frac{1}{i\sD _x+m_i}\delta (x,y)\Biggr] \gamma _5,
\label{11,51}
\end{eqnarray}

\begin{figure}[ht]
\begin{center}
\epsfig{file=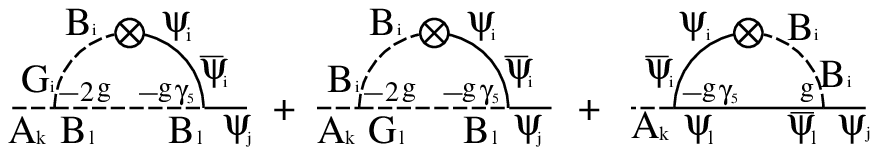,width=13cm}
\caption{Two Point Function $\langle \psi A\rangle _{tr}$(ii)}
\label{11,e}
\end{center}
\end{figure}

\begin{eqnarray}
& &\frac{\delta ^3\Gamma _1}
{\delta A_k(z)\delta \psi _j(y)\delta \bar{K}_A(x)}
\nonumber\\
&=&4g^2\sum _{il}\frac{m_i}{c_i}\Bigl[ 
\langle A_i(x)A_i(y)\rangle \langle \psi _i(x)\bar{\psi}_i(z)\rangle 
\langle \psi _l(z)\bar{\psi}_l(y)\rangle +\Bigl\{ \langle A_i(x)A_i(z)\rangle
\nonumber\\
& &\cdot \langle A_l(y)F_l(z)\rangle 
+\langle A_i(x)F_i(z)\rangle 
\langle A_l(y)A_l(z)\rangle \Bigr\} 
\langle \psi_i(x)\bar{\psi}_i(y)\rangle \Bigr] 
\nonumber\\
&=&-4ig^2\sum _{il}c_ic_lm_i\Biggl[ 
\frac{1}{i\sD _x+m_i}\delta (x,z)\cdot 
\frac{1}{i\sD _z+m_l}\delta (z,y)
\frac{1}{\Box _x+\xi {\cal R}-{m_i}^2}\delta (x,y)
\nonumber\\
& &-\frac{1}{\Box _x+\xi {\cal R}-{m_i}^2}\delta (x,z)\cdot 
\frac{m_i+m_l}{\Box _z+\xi {\cal R}-{m_l}^2}\delta (z,y)
\frac{1}{i\sD _x+m_i}\delta (x,y)\Biggr] ,
\label{11,53}
\end{eqnarray}

\begin{eqnarray}
& &\frac{\delta ^3\Gamma _1}
{\delta A_k(z)\delta \psi _j(y)\delta \bar{K}_B(x)}
\nonumber\\
&=&-4g^2\sum _{il}\frac{m_i}{c_i}\Bigl[ 
\langle B_i(x)B_i(y)\rangle \langle \psi _i(x)\bar{\psi}_i(z)\rangle 
\langle \psi _l(z)\bar{\psi}_l(y)\rangle -\Bigl\{ \langle B_i(x)B_i(z)\rangle
\nonumber\\
& &\cdot \langle B_l(y)G_l(z)\rangle 
+\langle B_i(x)G_i(z)\rangle 
\langle B_l(y)B_l(z)\rangle \Bigr\} 
\langle \psi_i(x)\bar{\psi}_i(y)\rangle \Bigr] \gamma _5
\nonumber\\
&=&4ig^2\sum _{il}c_ic_lm_i\Biggl[ 
\frac{1}{i\sD _x+m_i}\delta (x,z)\cdot 
\frac{1}{i\sD _z+m_l}\delta (z,y)
\frac{1}{\Box _x+\xi {\cal R}-{m_i}^2}\delta (x,y)
\nonumber\\
& &-\frac{1}{\Box _x+\xi {\cal R}-{m_i}^2}\delta (x,z)\cdot 
\frac{m_i+m_l}{\Box _z+\xi {\cal R}-{m_l}^2}\delta (z,y)
\frac{1}{i\sD _x+m_i}\delta (x,y)\Biggr] \gamma ^5,
\label{11,55}
\end{eqnarray}

\begin{eqnarray}
& &\frac{\delta ^3\Gamma _1}
{\delta A_k(z)\delta \psi _j(y)\delta \bar{K'}_A(x)}
\nonumber\\
&=&4g^2\sum _{il}\frac{\xi}{c_i}\Bigl[ 
\langle A_i(x)A_i(y)\rangle \langle \psi _i(x)\bar{\psi}_i(z)\rangle 
\langle \psi _l(z)\bar{\psi}_l(y)\rangle +\Bigl\{ \langle A_i(x)A_i(z)\rangle
\nonumber\\
& &\cdot \langle A_l(y)F_l(z)\rangle 
+\langle A_i(x)F_i(z)\rangle 
\langle A_l(y)A_l(z)\rangle \Bigr\} 
\langle \psi_i(x)\bar{\psi}_i(y)\rangle \Bigr] 
\nonumber\\ 
&=&-4ig^2\sum _{il}c_ic_l\xi \Biggl[ 
\frac{1}{i\sD _x+m_i}\delta (x,z)\cdot 
\frac{1}{i\sD _z+m_l}\delta (z,y)
\cdot \frac{1}{\Box _x+\xi {\cal R}-{m_i}^2}\delta (x,y)
\nonumber\\
& &
-\frac{1}{\Box _x+\xi {\cal R}-{m_i}^2}\delta (x,z)\cdot 
\frac{m_i+m_l}{\Box _z+\xi {\cal R}-{m_l}^2}\delta (z,y)
\cdot \frac{1}{i\sD _x+m_i}\delta (x,y)\Biggr] ,
\label{11,57}
\end{eqnarray}
and
\begin{eqnarray}
& &\frac{\delta ^3\Gamma _1}
{\delta A_k(z)\delta \psi _j(y)\delta \bar{K'}_B(x)}
\nonumber\\
&=&-4g^2\sum _{il}\frac{\xi}{c_i}\Bigl[ 
\langle B_i(x)B_i(y)\rangle \langle \psi _i(x)\bar{\psi}_i(z)\rangle 
\langle \psi _l(z)\bar{\psi}_l(y)\rangle -\Bigl\{ \langle B_i(x)B_i(z)\rangle
\nonumber\\
& &\cdot \langle B_l(y)G_l(z)\rangle 
+\langle B_i(x)G_i(z)\rangle 
\langle B_l(y)B_l(z)\rangle \Bigr\} 
\langle \psi_i(x)\bar{\psi}_i(y)\rangle \Bigr] \gamma _5
\nonumber\\
&=&4ig^2\sum _{il}c_ic_l\xi \Biggl[ 
\frac{1}{i\sD _x+m_i}\delta (x,z)\cdot 
\frac{1}{i\sD _z+m_l}\delta (z,y)
\cdot \frac{1}{\Box _x+\xi {\cal R}-{m_i}^2}\delta (x,y)
\nonumber\\
& &
-\frac{1}{\Box _x+\xi {\cal R}-{m_i}^2}\delta (x,z)\cdot 
\frac{m_i+m_l}{\Box _z+\xi {\cal R}-{m_l}^2}\delta (z,y)
\cdot \frac{1}{i\sD _x+m_i}\delta (x,y)\Biggr] \gamma ^5.
\label{11,59}
\end{eqnarray}

A two point function with a composite operators$(\bar{K}_{A^2})$ is 
\begin{eqnarray}
& &\frac{\delta ^3 W_1}
{\delta J_A^{k'}(z')\delta \eta ^{j'}(y')\delta \bar{K}_{A^2}(x)}
\nonumber\\
&=&-ig\int d^4u\sum _{ijk}
\langle A_{k'}(z')\bar{\psi}_{j'}(y')A_i(x)A_j(x)\psi _k(x)
{\cal L}_{int}(u)\rangle 
\nonumber\\
&=&4ig^2\int d^4ue_u\sum _{jk}\Bigl[ 
\langle A_{k'}(z')A_{k'}(x)\rangle 
\langle \psi _{j'}(u)\bar{\psi}_{j'}(y')\rangle \langle A_j(x)A_j(u)\rangle
\nonumber\\
& &\cdot \langle \psi _k(x)\bar{\psi}_k(u)\rangle + 
\langle A_{k'}(z')A_{k'}(u)\rangle 
\langle \psi _{j'}(x)\bar{\psi}_{j'}(y')\rangle 
\langle A_j(x)A_j(u)\rangle
\nonumber\\ 
& &\cdot \langle A_k(x)F_k(u)\rangle +\cdots \Bigr] 
\nonumber\\
&=&4ig^2\int d^4ud^4ve_ue_v\sum _{jk}\Bigl[ \Bigl\{ 
\delta (x,v)\langle A_j(v)A_j(u)\rangle 
\langle \psi _k(v)\bar{\psi}_k(u)\rangle 
\nonumber\\
& & + \delta (x,v)\langle A_j(v)A_j(u)\rangle 
\langle A_k(v)F_k(u)\rangle \Bigr\} 
\langle A_{k'}(z')A_{k'}(v)\rangle 
\nonumber\\
& &\cdot \langle \psi _{j'}(u)\bar{\psi}_{j'}(y')\rangle +\cdots \Bigr] .
\label{11,60}
\end{eqnarray}

We obtain (Figure~\ref{11,f})
\begin{eqnarray}
& &\frac{\delta ^3\Gamma _1}{\delta A_k(z)\delta \psi _j(y)
\delta \bar{K}_{A^2}(x)}
\nonumber\\
&=&-4ig^2\sum_{il}\langle A_l(z)A_l(y)\rangle 
\Bigl[ \delta (x,z)\langle \psi _i(z)\bar{\psi}_i(y)\rangle 
+\delta (x,y)\langle A_i(y)F_i(z)\rangle \Bigr] 
\nonumber\\
&=&4ig^2\sum _{il}c_ic_l\frac{1}{\Box _x+_xi {\cal R}-{m_l}^2}\delta (z,y)
\Biggl[ \delta (x,z)\frac{1}{i\sD _z+m_i}\delta (z,y)
\nonumber\\
& &-\delta (x,y)\frac{m_i}{\Box _y+\xi {\cal R}-{m_i}^2}\delta (y,z)\Biggr] .
\label{11,62}
\end{eqnarray}

\begin{figure}[ht]
\begin{center}
\epsfig{file=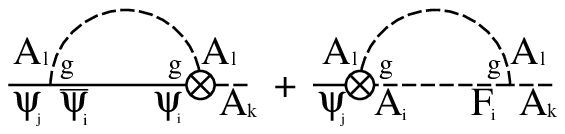,width=9cm}
\caption{Two Point Function $\langle \psi A\rangle _{tr}$(iii)}
\label{11,f}
\end{center}
\end{figure}

In the same way we obtain (Figure~\ref{11,g} and \ref{11,h})
\begin{eqnarray}
& &\frac{\delta ^3\Gamma _1}{\delta A_k(z)\delta \psi _j(y)
\delta \bar{K}_{B^2}(x)}
=4ig^2\sum_{il}\delta (x,y)\langle B_i(y)B_i(z)\rangle 
\langle B_l(y)B_l(z)\rangle 
\nonumber\\
& &\hspace{0.2cm}
=-4ig^2\sum _{il}c_ic_l\delta (x,y)\frac{m_i}{\Box _y+\xi {\cal R}-{m_i}^2}
\delta (y,z)\cdot \frac{1}{\Box _z+\xi {\cal R}-{m_l}^2}\delta (z,y),
\label{11,64}
\end{eqnarray}

\begin{figure}[ht]
\begin{center}
\epsfig{file=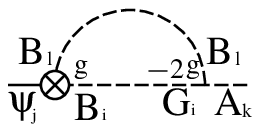,width=4cm}
\caption{Two Point Function $\langle \psi A\rangle _{tr}$(iv)}
\label{11,g}
\end{center}
\end{figure}

and
\begin{eqnarray}
& &\frac{\delta ^3\Gamma _1}{\delta A_k(z)\delta \psi _j(y)
\delta \bar{K}_{AB}(x)}
=4ig^2\sum _{il}\delta (x,z)\langle B_i(z)B_i(y)\rangle 
\langle \psi _l(z)\bar{\psi}_l(y)\rangle 
\gamma _5
\nonumber\\
& &\hspace{.3cm} =-4ig^2\sum _{il}c_ic_l\delta (x,z)
\frac{1}{\Box _z+\xi {\cal R}-{m_i}^2}\delta (z,y)
\cdot \frac{1}{i\sD _z+m_l}\delta (z,y)\gamma _5 .
\label{11,66}
\end{eqnarray} 

\begin{figure}[ht]
\begin{center}
\epsfig{file=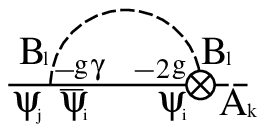,width=4cm}
\caption{Two Point Function $\langle \psi A\rangle _{tr}$(v)}
\label{11,h}
\end{center}
\end{figure}
Finally we compute the real scalar two point vertex function 
(Figure~\ref{11,i}) and spinor one (Figure~\ref{11,j})
\begin{eqnarray}
\frac{\delta ^2\Gamma _1}{\delta A_k(z)\delta A_j(x)}
&=&-2ig^2\sum_{il}c_ic_l\Biggl[ \tr{ 
\frac{1}{i\sD _x+m_i}\delta (x,z)\cdot 
\frac{1}{i\sD _z+m_l}\delta (z,x)}
\nonumber\\
& &-4\Biggl\{ 1+\frac{m_i^2}{\Box _x+\xi {\cal R}-{m_i}^2}\Biggr\} \delta (x,z)
\cdot \frac{1}{\Box _z+\xi {\cal R}-{m_l}^2}\delta (z,x)
\nonumber\\
& &-4\frac{m_i}{\Box _x+\xi {\cal R}-{m_i}^2}\delta (x,z)
\cdot \frac{m_l}{\Box _z+\xi {\cal R}-{m_l}^2}\delta (z,x)\Biggr] ,
\label{11,68}
\end{eqnarray}

\begin{figure}[ht]
\begin{center}
\epsfig{file=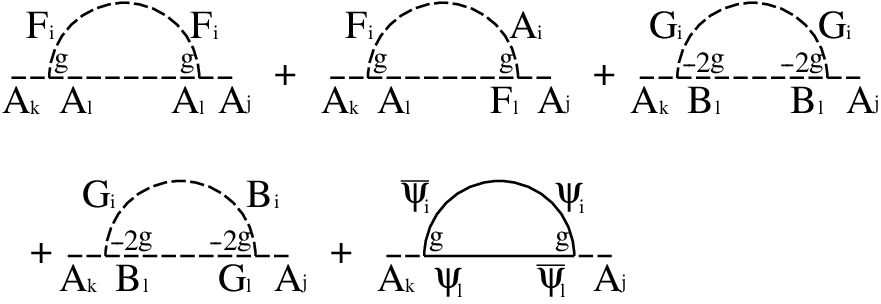,width=13cm}
\caption{Two Point Function $\langle AA\rangle _{tr}$}
\label{11,i}
\end{center}
\end{figure}

and
\begin{eqnarray}
\frac{\delta ^2\Gamma _1}{\delta \psi _j(y)\delta \bar{\psi}_k(x)}
&=&4ig^2\sum _{il}c_ic_l\frac{1}{\Box _x+\xi {\cal R}-{m_i}^2}\delta (x,y)
\Biggl[ \frac{1}{i\sD _x+m_l}\delta (x,y)
\nonumber\\
& &+\gamma _5\frac{1}{i\sD _x+m_l}\delta (x,y)\gamma _5
\Biggr] .
\label{11,70}
\end{eqnarray}

\begin{figure}[ht]
\begin{center}
\epsfig{file=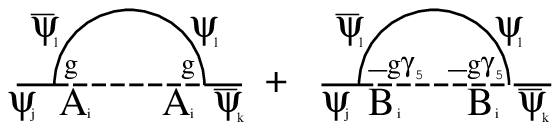,width=9cm}
\caption{Two Point Function $\langle \psi \psi \rangle _{tr}$}
\label{11,j}
\end{center}
\end{figure}

Inserting all the vertex functions into left hand side of 
Eq.$(\ref{11,46})$, we found the Ward-Takahashi 
identity for two-point functions satisfied. 
Other Ward-Takahashi 
identities for two-point functions which are obtained differentiating 
Eq.~(\ref{11,27}) with respect to $B_k(z)$, $F_k(z)$, or $G_k(z)$ 
instead of $A_k(z)$ can be checked in the same way.   

If we change the superpotential~$(\ref{11,0})$ as follows:
\begin{eqnarray}
P=\frac{1}{2}\sum _{\alpha}m_{\alpha}{\phi _{\alpha}}^2
+\frac{\sqrt{2}}{3}\sum _{\alpha \beta \gamma}
g_{\alpha \beta \gamma}\phi_{\alpha }\phi _{\beta}\phi _{\gamma} ,
\label{pgsp}
\end{eqnarray} 
then we obtain a Wess-Zumino model with many chiral multiplets which have 
various masses and coupling constants. Greek letters
indicate physical fields. 
It is easily found that SUSY is broken softly in these models as well
as the Wess-Zumino model with a chiral multiplet. 
Inclusion of their regulators is straightforward: 
\begin{eqnarray}
P=\frac{1}{2}\sum _i\sum _{\alpha} m_{\alpha i}{\phi _{\alpha i}}^2
+\frac{\sqrt{2}}{3}\sum _{ijk}\sum _{\alpha \beta \gamma }
g_{\alpha \beta \gamma }\phi_{\alpha i}\phi _{\beta j}\phi _{\gamma k}. 
\label{gsp}
\end{eqnarray} 
The subscriptions $\phi _{\alpha 0}$, $\phi _{\beta 0}$, $\phi _{\gamma 0}$ 
denote the physical fields and 
$\phi _{\alpha i}$, $\phi _{\beta j}$, $\phi _{\gamma k}$ where
$ijk\neq 0$, denote their regulators, respectively.

If we set the coupling constant $g_{122}=g_{121}=g_{221}=\frac{1}{2}g$ 
and others zero we obtain the same model with Wess-Zumino model with the 
flat direction which we shall study in Section~\ref{WZMFD(I)}.

Setting $gc_i \to g_{iij}c_i$ in the one point vertex functions and 
$g^2c_ic_l \to g_{ikl}g_{ijl}c_ic_l$ in the two point vertex functions 
we obtained above, we 
found that Pauli-Villars regularization preserves softly 
broken SUSY in the Wess-Zumino model (Eq.~(\ref{ssp})) with general
coupling ($g_{ijk}$).\footnote{In this case Pauli-Villars regularization 
does not regularize all the divergent diagrams.} 
Because the coupling in
Eq.~(\ref{gsp}) can be regarded as the special case of Eq.~(\ref{ssp}), 
Pauli-Villars regularization is suitable to the Wess-Zumino models
(Eq.~(\ref{pgsp})) in de Sitter space. 

%%%%%%%%%%%%%%%%%%%%%%%%%%%%%%%%%%%%%%%%%%%%%%%%%%%%%%%%%%%%%%%%%%%%%%%%%%%%%%%%%%%%%%% WESS-ZUMINO MODEL WITH FD (I) %%%%%%%%%%%%%%%%%%%%%%%%%%%%%%%%%%%%%%%%%%%%%%%%%%%%%%%%%%%%%%%%%%%%%%%%%%%%%%%%%%%%%%%%%%%%%%%%%%%%%%%%%%%%%%%%%%%%%

\section{Wess-Zumino Model with Flat Direction (I)}
\label{WZMFD(I)}

We compute a one-loop effective potential along the flat direction 
by using Pauli-Villars regularization. The simplest Wess-Zumino model that has
a flat direction is given by a Lagrangian with regulators:
\begin{eqnarray}
 e^{-1}{\cal L}&=&\sum _i\biggl[ 
\frac{1}{2c_{1i}}\Bigl\{ A_{1i}\Box A_{1i}+B_{1i}\Box B_{1i}
+i\bar{\psi}_{1i}\sD \psi _{1i}+{F_{1i}}^2+{G_{1i}}^2
\nonumber\\
& &+\xi _1{\cal R}({A_{1i}}^2+{B_{1i}}^2) 
+m_{1i}\bigl( 2A_{1i}F_{1i}-2B_{1i}G_{1i}
+\bar{\psi}_{1i}\psi _{1i}\bigr) \Bigr\} \nonumber\\
& &+\Bigl\{ 1\leftrightarrow 2\Bigr\} \Bigr] 
+g\sum_{ijk}\biggl[ (A_{1i}A_{2j}-B_{1i}B_{2j})F_{2k}
-(B_{1i}A_{2j}+A_{1i}B_{2j})G_{2k} 
\nonumber\\
& &+\frac{1}{2}\Bigl\{ (A_{2i}A_{2j}-B_{2i}B_{2j})F_{1k}
-2A_{2i}B_{2j}G_{1k}\Bigr\} 
+\bar{\psi}_{1i}(A_{2j}-\gamma _5B_{2j})\psi _{2k}\nonumber\\
& &+\frac{1}{2}\bar{\psi}_{2i}(A_{1j}-\gamma _5B_{1j})\psi _{2k}\biggr] .
\label{12,12}
\end{eqnarray}
Flat directions are 
\begin{eqnarray}
& &A_{1i},\, B_{1i}\neq 0\hspace{.5cm}\mbox{and}\hspace{.5cm} A_{2i}=B_{2i}=0,
\end{eqnarray}
where 
\begin{eqnarray}
& &F_{1i}=-m_{1i}A_{1i}-\frac{g}{2}\sum
_{jk}(A_{2j}A_{2k}-B_{2j}B_{2k})=0,\nonumber\\
& &G_{1i}= m_{1i}B_{1i}+g\sum _{jk}A_{2j}B_{2k}=0,\nonumber\\ 
& &F_{2i}=-m_{2i}A_{2i}-g\sum_{jk}(A_{1j}A_{2k}-B_{1j}B_{2k})=0,\nonumber\\ 
& &G_{2i}= m_{2i}B_{2i}+g\sum _{jk}(A_{1j}B_{2k}+A_{1k}B_{2j})=0,
\end{eqnarray} 
if $m$s and $\xi $s are all equal zero. 
The Pauli-Villars constraints are $\sum _ic_{1i}{m_{1i}}^p=0$, and 
$\sum _ic_{2i}{m_{2i}}^p=0$ for $p=0,1,2$.

We compute one and two point functions by using the regulators and
n-point functions $(n\geq 3)$ without regulators, since one and two point 
functions are divergent while others are convergent in a effective potential.
Using the same procedure of previous section we obtain the non-vanishing 
one point function:
\begin{eqnarray}
\frac{\delta \Gamma_1}{\delta A_{1j}(y)}=\frac{g}{2}\sum _i\Bigl[ 
\tr{\langle \psi _{2i}(y)\bar{\psi}_{2i}(y)\rangle }
-4\langle A_{2i}(y)F_{2i}(y)\rangle \Bigr] 
\label{12,14}.
\end{eqnarray}
We invoke the curvature expansion of propagators. 
The curvature expansion is sufficient for our calculations since 
the divergent parts are linear in curvature.  
Using propagators represented in momentum space in Eq.~$(\ref{11,10})$ and 
$(\ref{11,18})$ we obtain 
\begin{eqnarray}
\frac{\delta \Gamma _1}{\delta A_{1j}(y)}&=&
2ig\sum _ic_{2i}m_{2i}\int \frac{d^4k}{(2\pi )^4}\biggl( 
\frac{1}{4}-\xi _2\biggr) \frac{{\cal R}}{(k^2+{m_{2i}}^2)^2}
+{\cal O}({\cal R}^2)
\label{12,15}\\
&=&\frac{2g}{16\pi ^2}\sum _ic_{2i}m_{2i}\biggl( \frac{1}{4}-\xi _2\biggr) 
{\cal R}\ln {\frac{{m_{2i}}^2}{\mu ^2}}+{\cal O}({\cal R}^2),
\label{12,16}
\end{eqnarray}
where parameter $\mu $ is just introduced to make the argument of logarithm 
dimensionless while it does not contribute to the result at all because of 
the Pauli-Villars constraint.

Next we calculate a two point function: 
\begin{eqnarray}
\frac{\delta ^2\Gamma _1}{\delta A_{1k}(z)\delta A_{1j}(y)}&=&
-\frac{i}{2}g^2\sum _{il}\Bigl[ 4\langle A_{2i}(z)A_{2i}(y)\rangle 
\langle F_{2l}(z)F_{2l}(y)\rangle 
\nonumber\\
& &+4\langle A_{2i}(z)F_{2i}(y)\rangle \langle F_{2l}(z)A_{2l}(y)\rangle 
\nonumber\\
& &-\tr{\langle \psi _{2i}(z)\bar{\psi}_{2i}(y)\rangle 
\langle \psi _{2l}(y)\bar{\psi}_{2l}(z)\rangle }
\Bigr] .
\label{12,18}
\end{eqnarray}
Using Riemann normal coordinates around $x$.
\begin{eqnarray}
& &\frac{\delta ^2\Gamma _1}{\delta A_{1k}(z)\delta A_{1i}(x)}=
2ig^2\sum _{jl}c_{2j}c_{2l}\int \frac{d^4k}{(2\pi)^4}
\frac{d^4p}{(2\pi)^4}e^{i(k-p)\cdot (z-x)}\Biggl[ 
-\frac{1}{k^2+{m_{2j}}^2}
 \nonumber\\
& &\hspace{1cm}+\frac{2(k\cdot p)+{m_{2j}}^2
+{m_{2l}}^2}{2(k^2+{m_{2j}}^2)(p^2+{m_{2l}}^2)}
+\frac{(1-3\xi _2) {\cal R}}{3(k^2+{m_{2j}}^2)^2}
-\frac{2{\cal R}_{ab}k^ak^b}{3(k^2+{m_{2j}}^2)^3}
\nonumber\\
& &\hspace{1cm}+{\cal R}\Biggl\{ 
-\frac{4(1-3\xi _2)(m_{2j}+m_{2l})^2+5(k\cdot p)
-2m_{2j}m_{2k}}{12(k^2+{m_{2j}}^2)(p^2+m_{2l}^2)^2}
\nonumber\\
& &\hspace{1cm}+\frac{p^2}{6} \frac{2(k\cdot p)+{m_{2j}}^2
+{m_{2l}}^2}{(k^2+{m_{2j}}^2)(p^2+{m_{2l}}^2)^3}\Biggr\} 
+{\cal O}({\cal R}^2)\Biggr] .
\label{12,19}
\end{eqnarray}
Setting $k-p=q$ and performing volume integration of $z$ we obtain
\begin{eqnarray}
& &\int d^4ze_z\frac{\delta ^2\Gamma _1}{\delta A_{1k}(z)\delta A_{1i}(x)}
\nonumber\\
&&=2ig^2\sum _{jl}c_{2j}c_{2l}\int \frac{d^4k}{(2\pi)^4}{\cal R}
\Biggl[ \frac{(1-4\xi _2)}{4(k^2+{m_{2j}}^2)^2}
+\frac{{m_{2j}}^2}{6(k^2+{m_{2j}}^2)^3}
\nonumber\\
& &\hspace{.4cm}-\frac{6(1-4\xi _2)m_{2j}m_{2l}+3(1-4\xi _2){m_{2j}}^2
+2(1-6\xi _2){m_{2l}}^2}{12(k^2+{m_{2j}}^2)(k^2+{m_{2l}}^2)^2} 
\Biggr] +{\cal O}({\cal R}^2) \nonumber\\
&&=\frac{2g^2}{16\pi ^2}\biggl( \frac{1}{4}-\xi _2\biggr) {\cal R}
\sum _{jl}c_{2j}c_{2l}\Biggl[ \ln{\frac{m_{2j}m_{2l}}{\mu ^2}}
+\frac{m_{2j}+m_{2l}}{m_{2j}-m_{2l}}\ln {\frac{m_{2j}}{m_{2l}}}\Biggr] 
+{\cal O}({\cal R}^2) .
\label{12,21}
\end{eqnarray} 
At the fianl step we used the Pauli-Villars constraint 
$\sum _jc_{2j}=0$. 
An effective action can be expanded in a field
\begin{eqnarray}
\Gamma [\phi (x)]&=&\Gamma [0]+\int d^4xe_x
\left. \frac{\delta \Gamma }{\delta \phi (x)}\right| _{\phi =0}\phi (x)
\nonumber\\
& &+\frac{1}{2!}\int d^4xd^4ye_xe_y
\left. \frac{\delta ^2\Gamma }
{\delta \phi (x)\delta \phi (y)}\right| _{\phi =0}
\phi (x)\phi (y)+\cdots
\label{12,22}
\end{eqnarray}
and for a constant field we obtain an effecitve potential:
\begin{eqnarray}
-V(\phi)=-V(0)
+\left. \frac{\delta \Gamma }{\delta \phi (x)}\right| _{\phi =0}\phi  
+\frac{1}{2!}\int d^4ye_y\left. \frac{\delta ^2\Gamma }
{\delta \phi (x)\delta \phi (y)}\right| _{\phi =0}
\phi ^2+\cdots . 
\label{12,23}
\end{eqnarray}  
We set $m_{1i}=0$ to obtain exact flat 
directions in flat space limit. 
We take $\langle A_{10}\rangle =\phi $ and others are zero, since the
direction is a flat direction at tree level if $\xi_1$ vanishes. 
The effective potential along the direction is written as follows
\begin{eqnarray}
V_{eff}(\phi )&=&-\frac{1}{\sqrt{2}}J{\cal R}\phi 
-\frac{1}{2}\xi _1{\cal R}\phi ^2
+\frac{2g}{16\pi ^2}\sum _ic_{2i}m_{2i}\biggl( \frac{1}{4}-\xi _2\biggr) 
{\cal R}\ln{\frac{{m_{2i}}^2}{\mu ^2}}\cdot \phi 
\nonumber\\
& &+\frac{g^2}{16\pi ^2}\sum _{ij}c_{2i}c_{2j}
\biggl( \frac{1}{4}-\xi _2\biggr) {\cal R}\Biggl[ 
\ln{\frac{m_{2i}m_{2j}}{\mu ^2}}+\frac{m_{2i}+m_{2j}}{m_{2i}-m_{2j}}
\ln{\frac{m_{2i}}{m_{2j}}}\Biggr] \phi ^2
\nonumber\\
& &+\biggl[f(\phi)-f'(0)\phi 
-\frac{1}{2!}f''(0)\phi ^2\biggr] {\cal R} +{\cal O}({\cal R}^2),
\label{12,24} 
\end{eqnarray} 
where the term $-J{\cal R}\phi /\sqrt{2}$ is introduced for renormalization of 
linear term. The linear term does not 
affect one-loop calculation. 
The last line in Eq.~$(\ref{12,24})$ corresponds to the interaction 
part$(n\geq 3)$ which are all convergent. The function $f(\phi )$
is derived in the next section.
We renormalize one and two point vertex function at flat space limit 
as follows:
\begin{eqnarray}
\left.\frac{\partial V_{eff}}{\partial \phi }\right|_{\phi =M}
&=&-\frac{1}{\sqrt{2}}J{\cal R}-\xi _1{\cal R}M
+\frac{2g}{16\pi ^2}\sum _ic_{2i}m_{2i}\biggl( \frac{1}{4}-\xi _2\biggr) 
{\cal R}\ln{\frac{{m_{2i}}^2}{\mu ^2}}
\nonumber\\
& &+\frac{2g^2}{16\pi ^2}\sum _{ij}c_{2i}c_{2j}
\biggl( \frac{1}{4}-\xi _2\biggr) {\cal R}\Biggl[ 
\ln{\frac{m_{2i}m_{2j}}{\mu ^2}}+\frac{m_{2i}+m_{2j}}{m_{2i}-m_{2j}}
\ln{\frac{m_{2i}}{m_{2j}}}\Biggr] M 
\nonumber\\
& &+\biggl[f'(M)-f'(0)-f''(0)M \biggr] {\cal R}
\label{12,26}\\
&=&-\frac{1}{\sqrt{2}}J_R{\cal R}-\xi _{1R}{\cal R}M,
\label{12,27}\\
\left.\frac{\partial ^2V_{eff}}{\partial \phi ^2}\right|_{\phi =M}
&=&-\xi _1{\cal R}+\frac{2g^2}{16\pi ^2}\sum _{ij}c_{2i}c_{2j}
\biggl( \frac{1}{4}-\xi _2\biggr) {\cal R}\Biggl[ 
\ln{\frac{m_{2i}m_{2j}}{\mu ^2}}+\frac{m_{2i}+m_{2j}}{m_{2i}-m_{2j}}
\ln{\frac{m_{2i}}{m_{2j}}}\Biggr] 
\nonumber\\
& &+\biggl[f''(M)-f''(0) \biggr] {\cal R}
\label{12,28}\\
&=&-\xi _{1R}{\cal R} .
\label{12,29}
\end{eqnarray}
The effective potential is written in terms of renormalized parameters 
as follows
\begin{eqnarray}
V_{eff}(\phi )=-\frac{1}{\sqrt{2}} J_R{\cal R}\phi 
-\frac{1}{2}\xi _{1R}{\cal R}\phi ^2
+V_{one-loop}(\phi)+V_{counter}(\phi),
\label{12,8}
\end{eqnarray}
where
\begin{eqnarray}
V_{counter}(\phi )&\equiv &\frac{1}{\sqrt{2}}(-J+J_R){\cal R}\phi 
-\frac{1}{2}(\xi _1-\xi _{1R}){\cal R}\phi ^2.
\label{12,30}
\end{eqnarray}
Finally we obtain the effective potential written in terms of renormalized 
parameters defined by Eq.$(\ref{12,27})$ and $(\ref{12,29})$ as follows
\begin{eqnarray}
V_{eff}(\phi )&=&\biggl[ -\frac{1}{\sqrt{2}}J\phi 
-\frac{1}{2}\xi _1\phi ^2 +f(\phi)-f'(M)(\phi -M)
-\frac{1}{2}f''(M)(\phi -M)^2\biggr] {\cal R} 
\nonumber\\
& & +{\cal O}({\cal R}^2) 
\label{12,31},
\end{eqnarray} 
where we have omitted the subscript (R).
%%%%%%%%%%%%%%%%%%%%%%%%%%%%%%%%%%%%%%%%%%%%%%%%%%%%%%%%%%%%%%%%%%%%%%%%%%%%%%%%%%%%%% WESS-ZUMINO MODEL WITH FD (II)%%%%%%%%%%%%%%%%%%%%%%%%%%%%%%%%%%%%%%%%%%%%%%%%%%%%%%%%%%%%%%%%%%%%%%%%%%%%%%%%%%%%%%%%%%%%%%%%%%%%%%%%%%%%%%%%%%%%%%%%

\section{Wess-Zumino Model with Flat Direction (II)}
\label{WZMFD(II)}
In this section we calculate the effective potential by proper-time
cutoff regularization. This regularization is exact in curvature, so it
is appricable to the very early universe.    
Eliminating the auxiliary fields in the Lagrangian 
(Eq.~$(\ref{12,12})$) without regulators, 
and setting $m_1=0$ we obtain the corresponding 
Euclidean potential: 
\footnote{The Euclidean coordinate $x^4=ix^0$; 
the Euclidean gamma matrices $\gamma _{E4}=\gamma _0$, 
$\gamma _{Ei}=i\gamma _i$, and $\gamma _E^5=-i\gamma _{E1}\gamma _{E2}
\gamma _{E3}\gamma _{E4}$. In our convention the form of 
bosonic part of Lagrangian is 
same as the Lorentzian one, so is potantial. 
We shall omit the subscript $E$ in the 
following.} 
\begin{eqnarray}
V&=&\frac{1}{2}\xi _1{\cal R}({A_1}^2+{B_1}^2) 
+\frac{1}{2}(\xi _2{\cal R}-m_2^2)({A_2}^2+{B_2}^2) 
-gm_2A_1({A_2}^2+{B_2}^2)\nonumber\\
& &-\frac{g^2}{8}({A_2}^2+{B_2}^2)^2
-\frac{g^2}{2}({A_1}^2+{B_1}^2)({A_2}^2+{B_2}^2) .
\end{eqnarray}
In Euclidean de Sitter space ($S^4$) ${\cal R}=-12a^{-2}$, where $a=H^{-1}$.
We decompose the scalar fields around the flat direction 
where $\langle A_1\rangle =\phi$ as 
\begin{eqnarray}
\phi _1=\frac{\phi +A_1+iB_1}{\sqrt{2}},\; 
\phi _2=\frac{A_2+iB_2}{\sqrt{2}} .
\end{eqnarray}
We obtain the Lagrangian of the fluctuation of quadratic order: 
\begin{eqnarray}
e^{-1}{\cal L}_2&=&-\frac{1}{2}\biggl[ A_1(\Box +\xi _1{\cal R})A_1
 +B_1(\Box +\xi _1{\cal R})B_1 -\bar{\psi _1}\sD \psi _1 \nonumber\\
& &+A_2\Bigl\{ \Box +\xi _2{\cal R}-(m_2+g\phi )^2\Bigr\} A_2
 +B_2\Bigl\{ \Box +\xi _2{\cal R}-(m_2+g\phi )^2\Bigr\} B_2
\nonumber\\
& &-\bar{\psi _2}\Bigl\{ \sD -(m_2+g\phi)\Bigr\} \psi _2 \biggr] ,
\end{eqnarray}
and we can read off the effective mass of the fluctuations that depends 
on the expectation value $\phi $. The result is summarized in 
Table~\ref{3,aa}. 
\begin{table}
\begin{center}
\renewcommand{\arraystretch}{1.4}
\begin{tabular}{ccc} \hline
 & $L$ & $m^2a^2$ \\ \hline
$A_1,\; B_1$ & $0$ & $12\xi _1$\\
$A_2,\; B_2$ & $0$ & $12\xi _2+(m_2+g\phi )^2a^2$ \\
$\psi _1$ & $1/2$ & $0$ \\
$\psi _2$ & $1/2$ & $(m_2+g\phi )^2a^2$ \\
\hline
\end{tabular}
\end{center}
\caption{ The mass spectrum of the fluctuations with the presence of the 
vacuum expectation value $\phi $.}
\label{3,aa}
\end{table}  
From the mass generating pattern in Table~\ref{3,aa} 
and from Appendix~\ref{pt}
we can readily obtain the effective potential.
Because the finite part of the effective potential, $\zeta '(0)$, 
cannot be expressed by elementary functions, we must invoke numerical 
integrations. The analytic evaluation is possible at flat space 
limit\footnote{More precisely, at the limit, $g^2\phi ^2a^2\to \infty$ and 
${m_2}^2a^2\to \infty$.}  
 From the general rule in Table~\ref{2,d} in Appendix~\ref{pt}, we have   
\begin{eqnarray}
V_{eff}(\phi)&=&\Biggl[ -\frac{1}{\sqrt{2}}Js\phi 
-\frac{1}{2}\xi _1\phi ^2
+\frac{1-4\xi _2}{64\pi ^2}(g\phi +m_2)^2\Biggl\{ 
\gamma -1-\ln {\frac{\Lambda ^2}{(g\phi +m_2)^2}}\Biggr\} \Biggr] {\cal R}
\nonumber\\
& &+{\cal O}({\cal R}^2) ,
\label{12,3}
\end{eqnarray}
where we recover ${\cal R}$ instead of $-12a^{-2}$.
 We can see that one and two point vertex 
functions are all the divergent parts.
We adopt the same renormaliztion condition as Eq.~(\ref{12,27}) 
and (\ref{12,29}).
Finally we obtain the explicit form of effective potential 
\begin{eqnarray}
V_{eff}(\phi)&=&-\frac{1}{\sqrt{2}}J{\cal R}\phi 
-\frac{1}{2}\xi _1{\cal R}\phi ^2 
+\frac{1}{16\pi ^2}\biggl(\frac{1}{4}-\xi _2\biggr) {\cal R}
\Biggl[ (g\phi +m_2)^2
\nonumber\\
& &\cdot \Biggl\{ 
\ln{\Biggl( \frac{g\phi +m_2}{gM +m_2}\Biggr) ^2}-1\Biggr\} 
-2(g\phi -gM)^2\Biggr] +{\cal O}({\cal R}^2)
\label{12,11}
\end{eqnarray}
up to constant. We omitted the subscript (R) in the above expression.
Since terms of $\phi ^n(n\geq 3)$ 
are independent of regularizations, we set
\begin{eqnarray}
f(\phi)=-\frac{1-4\xi _2}{64\pi ^2}(g\phi +m_2)^2
\ln \frac{\Lambda ^2}{(g\phi +m_2)^2}\label{12,32}
\end{eqnarray}
in Eq.~$(\ref{12,31})$. The right hand side of Eq.~$(\ref{12,32})$ is
robbed from Eq.~$(\ref{12,3})$. We then find that the renormalized 
effective potential $(\ref{12,31})$ is the same as one in 
Eq.~(\ref{12,11}) up to ${\cal O}({\cal R}^2)$ terms. This fact means that
they give the same effective potential at all orders of the
curvature, since difference between the two regularizations is to linear 
order in ${\cal R}$.\footnote{From the above 
argument we deduced that we can naively obtain the same effective potential 
(\ref{12,11}) by means of any possible 
regularization. Accordingly, our effective 
potential can possess merits of all possible regularizations.} 
Quadratic and higher order terms in ${\cal R}$ are convergent. 

\begin{figure}
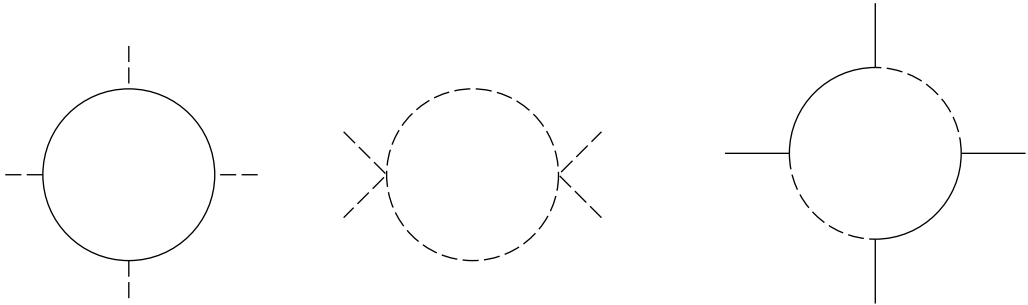

\begin{center}
\epsfig{file=4point.eps,width=8cm}
\hspace{1.5cm}\epsfig{file=f4point.eps,width=4cm}
\end{center}
\caption{The left two is the contributions to scalar four point 
vertex function where
  the solid line denotes a spinor field and the dashed line denotes a
  scalar field. The remaining is the contribution to spinor 
four point function.}
\label{4point} 
\end{figure}

Note that using proper-time cutoff regularization, 
the effective action of fermions can not be generated. 
By means of it we can regularize 
only the one-loop contribution along which a kind of field runs while the
effective action of fermions are made from the one-loop of several kinds 
of field. An example with a scalar self-coupling and the Yukawa coupling 
is depicted in Fig.~(\ref{4point}).   
Thus we cannot check the Ward-Takahashi identity for softly broken
SUSY by using proper-time cutoff regularization.

\begin{figure}
\begin{center}
\epsfig{file=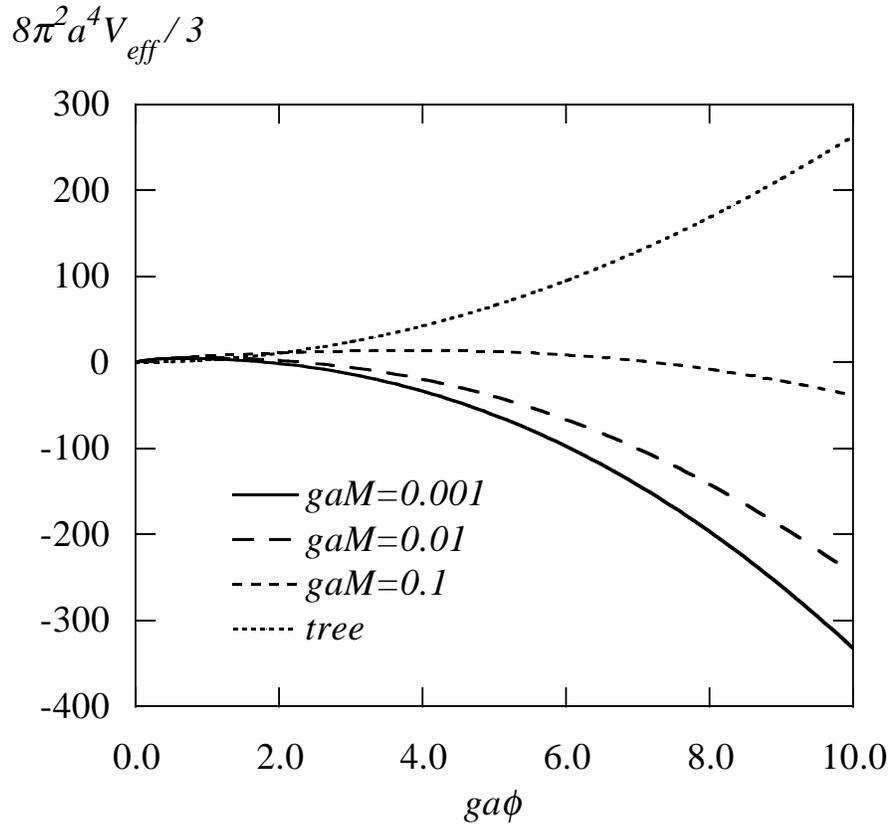,width=12cm}
\end{center}
\caption{The one-loop effective potential of the Wess-Zumino model with 
$g=0.1,\; J=0,\; 12\xi _i=0.001$. We have changed only the 
renormalization point as $gaM=0.001,0.01,0.1$ with all the other 
parameters fixed. 
The tree potential is also depicted for a 
comparison. We fixed $V_{eff}(0)=0$.}
\label{wz3} 
\end{figure}

\begin{figure}
\begin{center}
\epsfig{file=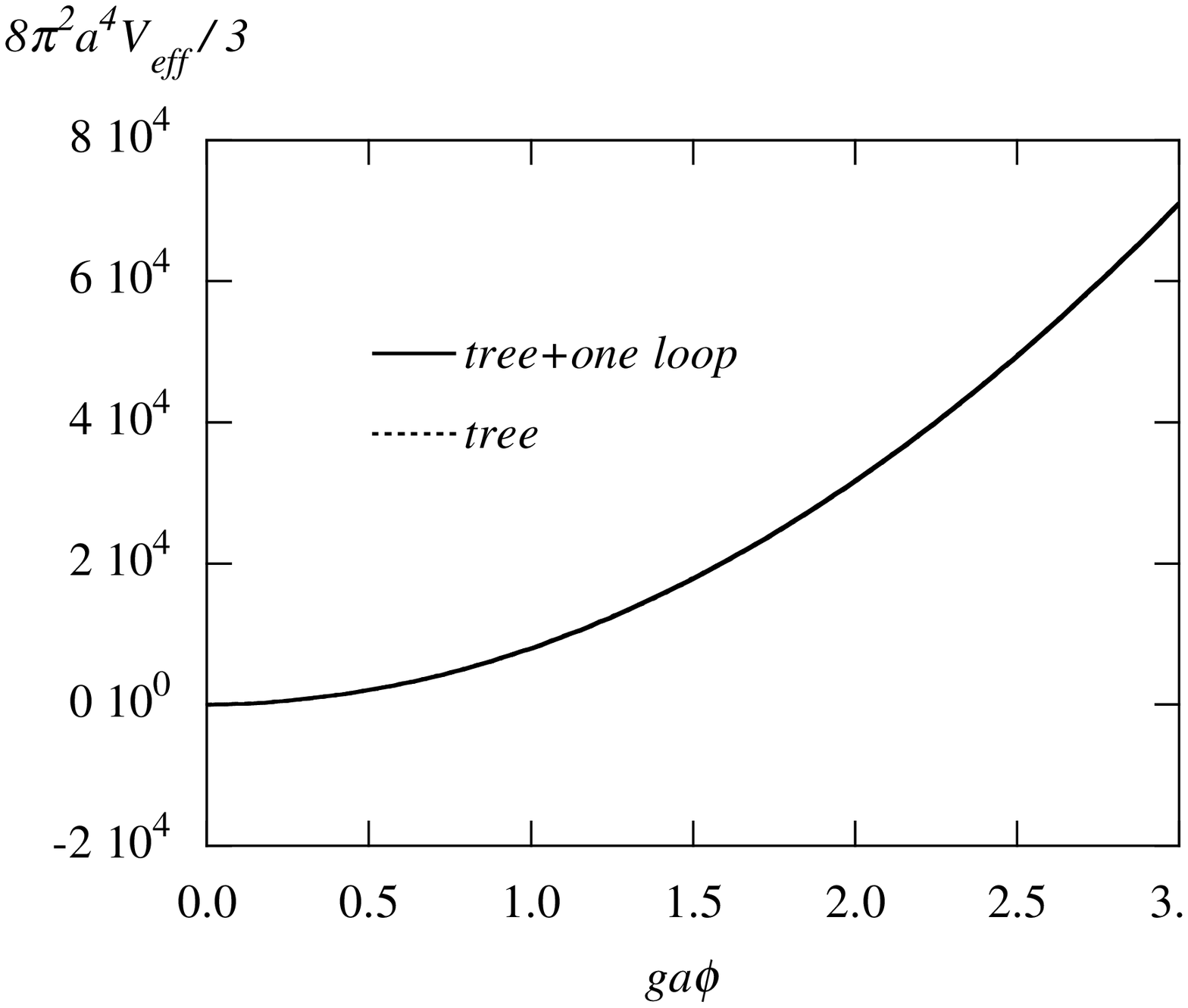,width=12cm}
\end{center}
\caption{The one-loop effective potential of the Wess-Zumino
  model with 
$g=0.1,\; J=0,\; \xi _i=1/4$. The tree potential is also depicted 
for a comparison.}
\label{wz1}
\end{figure}

Since $\xi _{\alpha}$ represent the strength of a coupling with (classical)
gravity, it is plausible to assume $\xi _1=\xi _2=\xi$. 
For the flat direction, which has $J\simeq 0$ and $\xi \simeq 0$, the
effective potential becomes 
\begin{eqnarray}
V_{eff}(\phi )\sim \frac{{\cal R}}{64\pi ^2}\Biggl[ (g\phi +m_2)^2
\Biggl\{ 
\ln{\Biggl( \frac{g\phi +m_2}{gM +m_2}\Biggr) ^2}-1\Biggr\} 
-2(g\phi -gM)^2\Biggr] +{\cal O}({\cal R}^2).
\end{eqnarray}
In the large $\phi $ region the asymptotic form of the effective potential 
is 
\begin{eqnarray}
V_{eff}(\phi )\to \frac{g^2}{64\pi ^2}{\cal R}\phi ^2
\ln{\Biggl( \frac{\phi }{M}\Biggr) ^2},
\end{eqnarray}
where ${\cal R}=-12H^2$. 
The effective potential becomes unbounded from below in the large $\phi$ 
region. This behavior is ture also 
in the large curvature region (Figure~(\ref{wz3})). 
This is because there is $\xi $ independent term in the
effective potential. Since one-loop approximation is reliable so long as 
\begin{eqnarray}
\frac{g^2}{64\pi ^2}\ln{\Biggl( \frac{\phi }{M}\Biggr) ^2}\aleq {\cal O}(1),
\end{eqnarray}
the effective 
curvature coupling constant could effectively be order unity. Accodingly, 
the flat direction is no longer flat.
If $J$ or $\xi$ is not so small, tree level potential is
dominant. In this case no flat direction exists even at tree level
(Figure~(\ref{wz1})).
  
%%%%%%%%%%%%%%%%%%%%%%%%%%%%%%%%%%%%%%%%%%%%%%%%%%%%%%%%%%%%%%%%%%%%%%%%%%%%%%%%%%%%%%%%%%%%%%%%% CONCLUSIONS AND DISCUSSION %%%%%%%%%%%%%%%%%%%%%%%%%%%%%%%%%%%%%%%%%%%%%%%%%%%%%%%%%%%%%%%%%%%%%%%%%%%%%%%%%%%%%%%%%%%%%%%%%%%%%%%%%%%%%%

\section{Conclusions and Discussion} 
\label{CD}

We have found that SUSY in curved
space is broken softly in the Wess-Zumino models. 
We consider that any suitable regularization must satisfy 
the Ward-Takahashi identity for softly broken SUSY. 
We found that Pauli-Villars regularization is a 
suitable regularization. 

Using the regularization we calculated the 
one-loop effective potential along a flat direction 
in de Sitter space. 
We calculated it to linear order in the
(space-time) curvature, or quadratic in the Hubble parameter, and found
that it is unbounded from below as follows:
\begin{eqnarray}
V_{eff}(\phi)\to 
-\frac{3g^2}{16\pi^2}H^2\phi ^2\ln \biggl( \frac{\phi ^2}{M^2}
\biggr).\label{40,1}
\end{eqnarray}
This result means that the effective potential is not flat at large $\phi $ 
region. 

The effective potential is exactly same as that is obtained by using 
proper-time cutoff regularization which is exact in curvature. 
The agreement of the two effective potentials to linear order 
in the curvature is sufficient,
since divergence and difference of regularizations appear only to the
order. 
From the above argument the effective potential we obtained 
possesses softly broken SUSY and reliability in the
large curvature region. 

The numerical calculation is depicted in Figure~$(\ref{wz3})$. 
It behaves like Eq.$(\ref{40,1})$ at the large $\phi $ region. 
The outstanding shape of the effective potential disappears as $H\to 0$. 
If the flat direction is
not exist at tree level, the potential does not receive much 
quantum correction (Figure~(\ref{wz1})).

The form of the effective potential is reliable 
even after inclusion of the renormalization group effect\cite{s&t}. 
 We argued that asymptotic behavior of the one-loop effective
potential is unavoidable irrespective of matter contents and 
detail of the model if the flat direction couples to a scalar 
multiplet\cite{s&t}.\footnote{Although the effective potential seems to
  be unbounded from below, (super)gravity shall bound the potential at 
the larger $\phi $ region.\cite{s&t}} 

Let us apply the unbounded potential to the Affleck-Dine mechanism.
 The scalar field begins to roll down along the 
unbounded potential in the inflationary era\cite{mt}. After inflation and
by the time when the Hubble parameter becomes comparable 
to the mass of the scalar
field, the potential becomes bounded from below. 
The scalar field stops rolling and has a large value along the 
direction. After that it rolls down to a true minimum (the origin). 
Accordingly, our effective potential will favor their mechanism 
though it is not always flat.\footnote{In the future work we will discuss the 
application to the Affleck-Dine mechanism in detail.}

%%%%%%%%%%%%%%%%%%%%%%%%%%%%%%%%%%%%%%%%%%%%%%%%%%%%%%%%%%%%%%%%%%%%%%%%%%%%%%%%%%%%%%%%%%%%%%%%% ACKNOWLEDGEMENT %%%%%%%%%%%%%%%%%%%%%%%%%%%%%%%%%%%%%%%%%%%%%%%%%%%%%%%%%%%%%%%%%%%%%%%%%%%%%%%%%%%%%%%%%%%%%%%%%%%%%%%%%%%%%%%%%%%%%%%%%
\section{Acknowledgement}

The author is grateful to H. Murayama, H. Suzuki, T. Yanagida, 
and J. Yokoyama. He also thanks K. Hikasa, M. Hotta, H. Inoue, K. Tobe, 
S. Watamura, M. Yamaguchi, and M. Yoshimura for useful comments.
\appendix
%%%%%%%%%%%%%%%%%%%%%%%%%%%%%%%%%%%%%%%%%%%%%%%%%%%%%%%%%%%%%%%%%%%%%%%%%%%%%%%%%%%%%%%%%%%%%%%%% NOTATIONS %%%%%%%%%%%%%%%%%%%%%%%%%%%%%%%%%%%%%%%%%%%%%%%%%%%%%%%%%%%%%%%%%%%%%%%%%%%%%%%%%%%%%%%%%%%%%%%%%%%%%%%%%%%%%%%%%%%%%%%%%%%%%%%%%

\section{Notations}
\label{N}

In this paper we use the convention as follows:
\begin{eqnarray}
& &\eta _{\mu\nu}=diag(-+++)\\
& &e\equiv \sqrt{-g}=\sqrt{-\mid g_{\mu\nu}\mid }\\
& &{\cal R}^{\rho}_{\; \, \sigma\nu\mu}\equiv \partial _{\mu}
\Gamma ^{\rho}_{\; \nu\sigma}
-\partial _{\nu}\Gamma ^{\rho}_{\; \mu\sigma}
+\Gamma ^{\lambda}_{\; \nu\sigma}
\Gamma ^{\rho}_{\; \mu\lambda}-\Gamma ^{\lambda}_{\; \mu\sigma}
\Gamma ^{\rho}_{\; \nu\lambda}
={\cal R}_{\mu\nu \sigma}^{\hspace{6mm}\rho}\\
& &{\cal R}_{\nu\mu}\equiv {\cal R}^{\rho}_{\; \, \nu\rho\mu}\\
& &{\cal R}\equiv {\cal R}^{\nu}_{\nu}.  
\end{eqnarray}
For the convetions of Gamma matrices are the same as those in Ref.~\cite{w&b}

%%%%%%%%%%%%%%%%%%%%%%%%%%%%%%%%%%%%%%%%%%%%%%%%%%%%%%%%%%%%%%%%%%%%%%%%%%%
%%%%%%%%%%%%%%%%%%%%%%%%%%%%%%%%%%%%%%%%%%%%%%%%%%%%%%%%%%%%%%%%%%%%%%%%%
\section{Conformally transformed Wess-Zumino model}

\label{CTWZ}
 
We investigate another reason why SUSY is broken softly in de Sitter 
space.
De Sitter space can be written by a conformally flat metric 
$g_{\mu\nu}(x)=\{ \Omega (x)\} ^2\eta_{\mu\nu}$.\footnote{It is known that 
the conformal flat metric
do not cover the whole de Sitter manifold. So we need to attach the patches 
appropriately. But we ignore this subtlety here.} 
Defining the 
tilde fields as 
\begin{eqnarray}
& &\tilde{A}=\Omega (x)A, \hspace{1cm} \tilde{B}=\Omega (x)B
\nonumber\\
& &\tilde{F}=\{ \Omega (x)\} ^2F, \hspace{1cm} \tilde{G}=\{ \Omega (x)\} ^2G
\nonumber\\
& &\tilde{\psi}=\{ \Omega (x)\} ^{3/2}\psi ,
\nonumber\\
\end{eqnarray}
we obtain
\begin{eqnarray}
{\cal L}&=&-\frac{1}{2}\Bigl( \partial _{\mu}\tilde{A}
\cdot \partial ^{\mu}\tilde{A}
+\partial _{\mu}\tilde{B}\cdot \partial ^{\mu}\tilde{B}\Bigr) 
+\frac{i}{2}\bar{\tilde{\psi}}\spartial \tilde{\psi}
+\frac{1}{2}\Bigl( \tilde{F}^2+\tilde{G}^2\Bigr) \nonumber\\
& &+a^{-2}\Omega ^2(1-6\xi) \Bigl( \tilde{A}^2+\tilde{B}^2\Bigr) 
+m\Omega \biggl( \tilde{A}\tilde{F}-\tilde{B}\tilde{G}
+\frac{1}{2}\bar{\tilde{\psi}}\tilde{\psi} \biggr) \nonumber\\
& &+g \Bigl[ \bar{\tilde{\psi}}(\tilde{A}-\gamma _5\tilde{B})
\tilde{\psi} +\tilde{F}\tilde{A}^2-\tilde{F}\tilde{B}^2-2\tilde{G}\tilde{A}
\tilde{B}\Bigr] ,
 \label{111,1}
\end{eqnarray}   
$a$ is the radius of de Sitter space.
This Lagrangian is corresponding to 
the Wess-Zumino model with a {\it soft SUSY breaking 
mass term} which depends on coordinates. 

%%%%%%%%%%%%%%%%%%%%%%%%%%%%%%%%%%%%%%%%%%%%%%%%%%%%%%%%%%%%%%%%%%%%%%%%%%%%
%%%%%%%%%%%%%%%%%%%%%%%%%%%%%%%%%%%%%%%%%%%%%%%%%%%%%%%%%%%%%%%%%%%%%%%%%%%
%%%%%%%%%%%%%%%%%%%%%%%%%%%%%%%%%%%%%%%%%%%%%%%%%%%%%%%%%%%%%%%%%%%%%%%%%%%%
\section{Proper-time cutoff}

\label{pt}

A one-loop effective potential is given by\cite{s&t,s} 
\begin{eqnarray}
V_{eff}(\phi)&=&V(\phi)+\frac{3}{16\pi ^2}\biggl \{ -\frac{1}{2}{\rm Res}
[\zeta (2)]\Lambda ^4-{\rm Res}[\zeta (1)]a^{-2}\Lambda ^2 \nonumber\\
& &+\zeta (0)\bigl[ \gamma -\ln (a\Lambda )^2\bigr] a^{-4}-\zeta '(0)a^{-4} 
\biggr \}.
\label{1,6}
\end{eqnarray}
$V(\phi)$ is a tree-level potential and $\zeta (n)$ is 
 generalized zeta function\cite{h} on $S^4$ evaluated by Allen 
$\cite{a}$. It is convenient to translate his results in terms of effective 
masses in the wave operator $\Delta =\, -\Box +m^2$. For 
spinorial fields $\Delta =\, (\sD -m)^{\dagger}
(\sD -m)$. Some relevant values are summarized in 
Table~\ref{2,b}. On the other hand, the derivative of zeta function at $s=0$, 
$\zeta '(0)$, is given by $\cite{a}$
\begin{eqnarray}
\zeta '(0)&=&-\frac{1}{3}(2L+1)\int _{0}^{y(L)}\biggl[ x^3+\Bigl( L+
\frac{1}{2}\Bigr) ^2x\biggr] \biggl[ \psi \Bigl( L+\frac{1}{2}+ix \Bigr)
\nonumber\\ 
& &+\psi \Bigl( L+\frac{1}{2}-ix \Bigr) \biggr] dx +c(L)
+\frac{2}{3}(2L+1)\biggl[ \zeta _R'\Bigl( -3,L+\frac{3}{2}\Bigr)\nonumber\\
& &-\Bigl(L+\frac{1}{2} \Bigr) ^2\zeta _R'\Bigl( -1,L+\frac{3}{2}\Bigr) 
\biggr] ,\label{1,7}
\end{eqnarray}
where constants $y(L)$ and $c(L)$ are given in Table~\ref{2,c}. In the above 
expression, $\psi (s)$ is the digamma function and $\zeta _R(s,\alpha )$ is 
the extended Riemann's zeta function. The integral in Eq.$(\ref{1,7})$ cannot 
be done analytically but we can evaluate the asymptotic behavior at $m^2a^2 
\, \to \, \infty$ by using the asymptotic form of the digamma function 
$\cite{a&s}$. We may also evaluate the integral Eq.$(\ref{1,7})$ numerically. 
In Table~\ref{2,d}, 
the asymptotic form of $\zeta '(0)$ at $m^2a^2 \, \to \, \infty$ 
are presented.
\begin{table}
\renewcommand{\arraystretch}{1.1}  
\begin{tabular}{lccc} \hline
 & 6$\times $Res[$\zeta $(2)] & 6$\times $Res[$\zeta $(1)] & 
180$\times \zeta $(0) \\ \hline
 real scalar & $1$ & $-m^2a^2+2$ & 
$15m^4a^4-60m^2a^2+58$ \\ 
Majorana spinor & $2$ & $-2m^2a^2-2$ &
 $30m^4a^4+60m^2a^2+11$ \\  
transverse massive vector & $3$ & $-3m^2a^2+3$ & 
$45m^4a^4-90m^2a^2-21$ \\ 
Rarita-Schwinger  & $4$ & $-4m^2a^2
-16$ & $60m^4a^4+480m^2a^2+802$ \\ 
symmetric transverse  & $5$ & $
-5m^2a^2-10$ & $ 75m^4a^4+300m^2a^2
-10$ \\
 traceless tensor & & & \\
\hline
\end{tabular}
\caption{ Some values of the generalized zeta function}
\label{2,b}
\end{table}    
\begin{table}
\begin{center}
\renewcommand{\arraystretch}{1.5}
\begin{tabular}{lccc} \hline
 & $L$ & $y(L)$ & $c(L)$ \\ \hline
 $\phi $ & $ 0 $ & $ \sqrt{m^2a^2-9/4}$ & $\frac{1}{12}m^4a^4
-\frac{7}{18}m^2a^2+\frac{29}{64}$ \\
 $\psi $ & $1/2$ & $|ma|$ & $\frac{1}{6}m^4a^4+\frac{1}{18}m^2a^2$ \\
 $A^{\mu}$ & $ 1 $ & $\sqrt{m^2a^2-13/4}$ & $\frac{1}{4}m^4a^4
-\frac{7}{6}m^2a^2+\frac{221}{192}$ \\
 $\psi ^{\mu}$ & $ 3/2$ & $|ma|$ & $\frac{1}{3}m^4a^4+\frac{13}{9}m^2
a^2$ \\
 $h^{\mu \nu}$ & $2$ & $\sqrt{m^2a^2-17/4}$ & $\frac{5}{12}m^4a^4
-\frac{5}{18}m^2a^2-\frac{3655}{576}$ \\
\hline
\end{tabular}
\end{center}
\caption{ Some constants appearing in Eq.~(\ref{1,7}) }
\label{2,c}
\end{table}

\begin{table}
\renewcommand{\arraystretch}{1.4}
\begin{tabular}{lc} \hline
 & 180$\times \zeta '(0)$ at $m^2a^2 \to \infty$ \\ \hline
 $\phi $ & $-15m^4a^4(\ln m^2a^2-3/2)+60m^2a^2(
\ln m^2a^2-1)-58\ln m^2a^2+\cdots $\\
 $\psi $ & $-30m^4a^4(\ln m^2a^2-3/2)-60m^2a^2(\ln 
m^2a^2-1)-11\ln m^2a^2+\cdots $\\
 $A^{\mu}$ & $-45m^4a^4(\ln m^2a^2-3/2)+90m^2a^2
(\ln m^2a^2-1)+21\ln m^2a^2+\cdots $\\
 $\psi ^{\mu}$ & $-60m^4a^4(\ln m^2a^2-3/2)-480m^2a^2
(\ln m^2a^2-1)-802\ln m^2a^2+\cdots $\\
 $h^{\mu \nu}$ & $75m^4a^4(\ln m^2a^2-3/2)-300m^2a^2(\ln m^2a^2-1)
+10\ln m^2a^2+\cdots $\\
\hline
\end{tabular}
\caption{The asymptotic forms of a derivative of the generalized zeta function 
at $s=0$}
\label{2,d}
\end{table}

\newpage

%%%%%%%%%%%%%%%%%%%%%%%%%%%%%%%%%%%%%%%%%%%%%%%%%%%%%%%%%%%%%%%%%%%%%%%%%%%%%%%%%%%%%%%%% THE BIBLIOGRAPHY %%%%%%%%%%%%%%%%%%%%%%%%%%%%%%%%%%%%%%%%%%%%%%%%%%%%%%%%%%%%%%%%%%%%%%%%%%%%%%%%%%%%%%%%%%%%%%%%%%%%%%%%%%%%%%%%%%%%%%%%%%%%%%%%%

\end{document}